\documentclass[11pt]{article}
\usepackage{graphicx}
\usepackage{amssymb,amsthm,amsmath}
\usepackage{mathtools}
\usepackage{comment}
\usepackage{xcolor,paralist,hyperref,titlesec,fancyhdr,etoolbox}
\usepackage{natbib}

\usepackage{url}
\usepackage{lipsum}
\usepackage{authblk}

\theoremstyle{definition}
\newtheorem{dfn}{Definition}
\newtheorem{lemma}{Lemma}
\newtheorem{prop}{Proposition}

\newtheorem{assumption}{Assumption}

\newcommand{\diff}{\mathrm{d}}
\newcommand{\Eq}[1]{{Eq.~(\ref{#1})}}
\newcommand{\defeq}{\coloneqq}

\newcommand{\lmb}{{\lambda_2}}
\newcommand{\de}{\delta}

\hypersetup{
	colorlinks=true,
	citecolor=blue,
	linkcolor=blue,
	urlcolor=blue,
}

\begin{document}

\title{Price Equilibria in a Spatial Competition with Captive Buyers\thanks{\scriptsize{The authors thank Tadashi Sekiguchi, Takatoshi Tabuchi, Noriaki Matsushima, Haruo Imai and the seminar participants at KIER (Institute of Economic Research, Kyoto University), AMES2024 (the 2024 Asia Meeting of the Econometric Society, East \& Southeast Asia), for useful discussions and suggestions. The previous version of this paper is available on SSRN: \url{https://ssrn.com/abstract=4870968}. The earlier version was entitled "Price Equilibria in a Spatial Competition with Uninformed Consumers".
% Declarations of interest: none. 
No funding was received for this work. All remaining errors are my own.}}} %%%%%%%%%%%%
\author[1]{Shinnosuke KAWAI\thanks{sskawai@shizuoka.ac.jp}}
\author[2]{Kuninori NAKAGAWA\thanks{nakagawak@em.u-hyogo.ac.jp}}

\affil[1]{\small{Department of Chemistry, Shizuoka University, Shizuoka, JAPAN 4228529.}}
\affil[2]{\small{School of Economics and Management, University of Hyogo, Kobe, JAPAN 6512197.}}

\date{\today}

% \keywords{keyword1, Keyword2, Keyword3, Keyword4}
% \begin{keyword}
% \kwd{Spatial competition}
% \kwd{Captive Buyers}
% \kwd{Linear transportation cost}
% \kwd{Hotelling's location model}
% \kwd{Varian's sales model}
% \end{keyword}

% \begin{keyword}[class=JEL] %% alphabetical order
% \kwd{D11}
% \kwd{D21}
% \kwd{D43}
% \kwd{D82}
% \end{keyword}

\maketitle 
% \maketitle

 %%%%%%%%%%

\begin{abstract}
This paper explores price competition with exogenous product differentiation in a spatial model similar to that of Nakagawa (2023). Nakagawa examines product differentiation within the framework of Varian (1980). Nakagawa integrates Varian's concept of uninformed consumers, who lack complete price information, into a spatial model based on Hotelling (1929). While Nakagawa placed informed consumers at the center of the Hotelling line and used quadratic transportation costs, our study employs a uniform distribution of informed consumers and linear transportation costs. This approach enables a more direct comparison with established spatial competition literature, particularly Osborne and Pitchik (1987).
We classify equilibrium candidates and characterize the parameter regions corresponding to each equilibrium. There is no pure equilibrium in the region where we construct mixed strategy equilibria. Furthermore, we compare the expected profit in the equilibrium of our model with the findings of Osborne and Pitchik (1987). Finally, we discuss the impact of captive buyers on the nature of spatial competition. 
\end{abstract} %%%%%%%%%

\section{Introduction}
According to the theory of perfect competition, temporary discounts (sales) do not occur in market equilibrium. \citet{v} showed that in price competition for homogeneous goods, a mixed strategy equilibrium emerges in the model where some consumers lack price information for some goods. These consumers are called uninformed consumers. It is well known that this equilibrium price dispersion violates the law of one price. This result relies heavily on the assumption of an uninformed consumer.

In Varian's model, the uninformed consumer knows only one firm's price information. Therefore, the word “uninformed" here means that the consumer has less price information than the informed consumer. The uninformed consumer makes decisions by observing only one firm's prices. In other words, “uninformed" can be interpreted as a situation where rivals' price information is blocked.  

Varian's model focuses on asymmetric price information for a homogeneous good. Even in the case of differentiated goods, some consumers may buy only from one firm without checking the other firm's goods. For example, consumers who prefer a particular clothing brand may purchase their favorite brand without checking any price information about the brand they are not interested in.

\citet{nakagawa} analyzed product differentiation between firms in Varian's model. He introduces explicit product differentiation into Varian's model, such as a simple spatial model of \citet{ho}. He analyzed product differentiation when, as in Varian's sales model, some consumers are uninformed about some prices. His model has room for improvement when viewed through the lens of Hotelling's one-dimensional spatial competition model.

First, \citeauthor{nakagawa} assumed informed consumers concentrated at $1/2$ in the interval $[0,1]$. Extending the distribution of consumers to a uniform distribution would allow us to consider the situation where the firms share the market of informed consumers. Second, \citeauthor{nakagawa} uses a quadratic transportation cost function.
\citet{da} demonstrates that transportation costs are crucial in Hotelling's model. For example, in Hotelling's one-dimensional model, linear transportation costs could lead to a winner-take-all scenario within the market of informed consumers.
 
The present paper considers a uniform distribution of informed consumers over the interval and linear transportation costs. 
This extension enables us to contrast our model with results from the standard literature on Hotelling's spatial competition model, e.g., \citeauthor{da}, \citet{op}, and \citet{https://doi.org/10.1111/jems.12032}. They provide an important study of price competition in the context of spatial competition.\footnote{To be more precise, they analyze free on board (FOB) pricing using a one-dimensional spatial competition model. The FOB price is also called the mill price.} Our model corresponds to the analysis of the model that adds uninformed consumers to the framework analyzed by \citeauthor{op}.

While the setting of the informed consumer follows Hotelling's original model, we adopt the same assumption of uninformed consumers as in Nakagawa's model. Our model assumes that uninformed consumers are located at both ends of the line, creating a market that an opponent cannot access. This market is always protected, leading us to interpret uninformed consumers' behavior as if they face prohibitively high transportation costs to the other end. Focusing on their behavior, we call them “captive" buyers.

We present equilibria in which both firms compete on price while taking into account the profits from captive buyers. We characterize all the pure equilibria. There is no pure equilibrium in the region where we analyze mixed strategy equilibria. In these mixed strategy equilibria, our analysis focused on the equilibrium where one firm could attract all the informed consumers, while the other only drew in captive buyers. Furthermore, we compare the expected profits achieved in our equilibrium with those found in \citeauthor{op}. We also discuss how competition changes when captive buyers exist at both ends of Hotelling's line. 

Our paper is organized as follows. In Section 2, we define the model. In Section 3, we present pure strategy equilibria. In Section 4, we present mixed strategy equilibria. Section 5 discusses the properties of price competition in our model. Section 6 is the conclusion.

\section{Model}
In our model, there are two firms, indexed as $i=1,2$. Both firms are located within the interval $[0,1]$, where $(z_1, z_2)$ represents the location points of the two firms. Given a pair of locations, the firms compete on price, $p_i$. We examine the price subgame that occurs after both firms simultaneously choose their locations $(z_1, z_2)$ within the interval $[0, 1]$.

\citeauthor{op} assume that consumers are fully informed. In our model, similar to Varian's model, we consider two types of consumers: informed and uninformed. Informed consumers, which will be denoted by $C_3$ hereafter, are uniformly distributed in the interval $(0,1)$. Uninformed consumers $C_1$ and $C_2$ are located at the endpoints, $0$ and $1$, respectively. Consumers $C_1$ at endpoint $0$ buy only Firm 1's product, while consumers $C_2$ at endpoint $1$ buy only Firm 2's product. This behavior can also be interpreted as loyalty to the seller, which may reflect the buyer's preferences. We refer to these types of consumers as “captive buyers." Additionally, we assume that each type of consumer has measure $1$.

We define each consumer's utility when they purchase a product with its characteristic $z_i$ at $p_i$ as follows. The reservation utilities for each type of consumer are set to $1$. Each type of consumer $C_k\:(k=1,2,3)$ purchases one unit of the product from either of the two firms. $t^{C_3}$ denotes each informed consumer's ideal point. Every informed consumer $t$ is assumed to be heterogeneous. In the Hotelling model, distance indicates a preference for proximity. All consumers pay a transportation cost for each unit of distance to a firm. We assumed linear transport costs. Every consumer chooses their action to maximize their utility:
\begin{align}
&u^{C_1}=1-(p_1+|z_1|),\\
&u^{C_2}=1-(p_2+|1-z_2|),\\
&u^{C_3}=1-\{p_i+|t^{C_3}-z_i|\}.
\end{align}
Informed consumers $C_3$ evaluate distance for all products. Utility functions of captive buyers $C_1$ and $C_2$, who only evaluate the distance from a specific product, will show a specific bias for preference. 

Furthermore, we assume that the reservation value of captive buyers for each firm's product is equal to that of informed consumers for both firms' products, and we normalize these reservation values to $1$.

\citeauthor{op} assume that consumers' reservation utility is sufficiently large to ensure that no consumers opt out of consuming a product.\footnote{\citet{ECONOMIDES1984345} discuss consumers' reservation utility in the usual model of spatial competition.} However, captive buyers exist in our model. If we assume that these buyers have unlimited reservation utility, firms can charge them any price. Therefore, we define a reservation utility for all consumers.  

Firm $i$'s profit is defined by the sum of the profit gained from both $C_i$ and $C_3$ consumers' market, given each consumer's choice for the firm's prices.
\begin{align}
\pi_1(p_1,p_2)=\pi_1^{C_1}(p_1)+\pi_1^{C_3}(p_1,p_2),\\
\pi_2(p_1,p_2)=\pi_2^{C_2}(p_2)+\pi_2^{C_3}(p_1,p_2).
\end{align}
Here, we assume that the production cost is zero and that firms' capacity constraints are not binding. The latter is the basic assumption of Bertrand's model of an oligopoly, cf. \citet{KS0af903d7-21a1-371d-9243-9ad71af18010}. 

Now we consider each type of consumer's choice and a firm's profit given the pair $(z_1,z_2)$. $\pi_i(p_1,p_2)$ is the total sum of $\pi_i^{C_i}(p_i)$ and $\pi_i^{C_3}(p_1,p_2)$. The former denotes Firm $ i$'s profit gained from $C_i$ because $C_j, j \ne i$ does not purchase the product from Firm $i$. The latter denotes Firm $ i$'s profit gained from $C_3$. First, we obtain $\pi_i^{C_i}(p_i), i=1,2$, as follows:
\begin{align}
\pi_1^{C_1}(p_1)=
\begin{dcases}
p_1, &\text{if $p_1 \le 1-z_1$,}\\
0, &\text{otherwise.}
\end{dcases}
\end{align}
\begin{align}
\pi_2^{C_2}(p_2)=
\begin{dcases}
p_2, &\text{if $p_2 \le z_2$,}\\
0, &\text{otherwise.}
\end{dcases}
\end{align}
Next we define $\pi_1^{C_3}(p_1,p_2)$. We obtain each informed consumer's utility as follows: 
\begin{align}
u^{C_3}=
\begin{dcases}
1-\{p_1+|t^{C_3}-z_1|\}, &\text{if they buy Firm 1's product,}\\
1-\{p_2+|t^{C_3}-z_2|\}, &\text{if they buy Firm 2's product,}\\
0, &\text{otherwise.}
\end{dcases}
\end{align}

Informed consumers buy the product with higher utility. Hereafter, we assume that $z_1$ is the left-hand side of $z$ and $z_1 \le z_2$. We define Firm $1$'s profit $\pi_1^{C_3}(p_1,p_2)$ as an expected profit $\mathrm{E}[\pi_1](p_1)$ when it chooses a price $p_1$. Let $F_2$ be the cumulative probability distribution of prices $p_2$ of Firm $2$. Thus, we have the following integration over the price distribution of its counterpart,
\begin{align}
& \mathrm{E}[\pi_1](p_1) = p_1 \left(\int_{p_1-\de}^{p_1+\de} \frac{ z_1+z_2-p_1+p_2}{2} \diff F_2^\star(p_2) + \left( 1 - F_2^\star(p_1+\de) \right)  \right). %\label{eq:expect1}
\end{align}
Here, $\delta := z_2-z_1$. We can define the expected profit of Firm $2$ in the same way.

\begin{figure}[ht]
\centering 
\includegraphics[width=6cm]{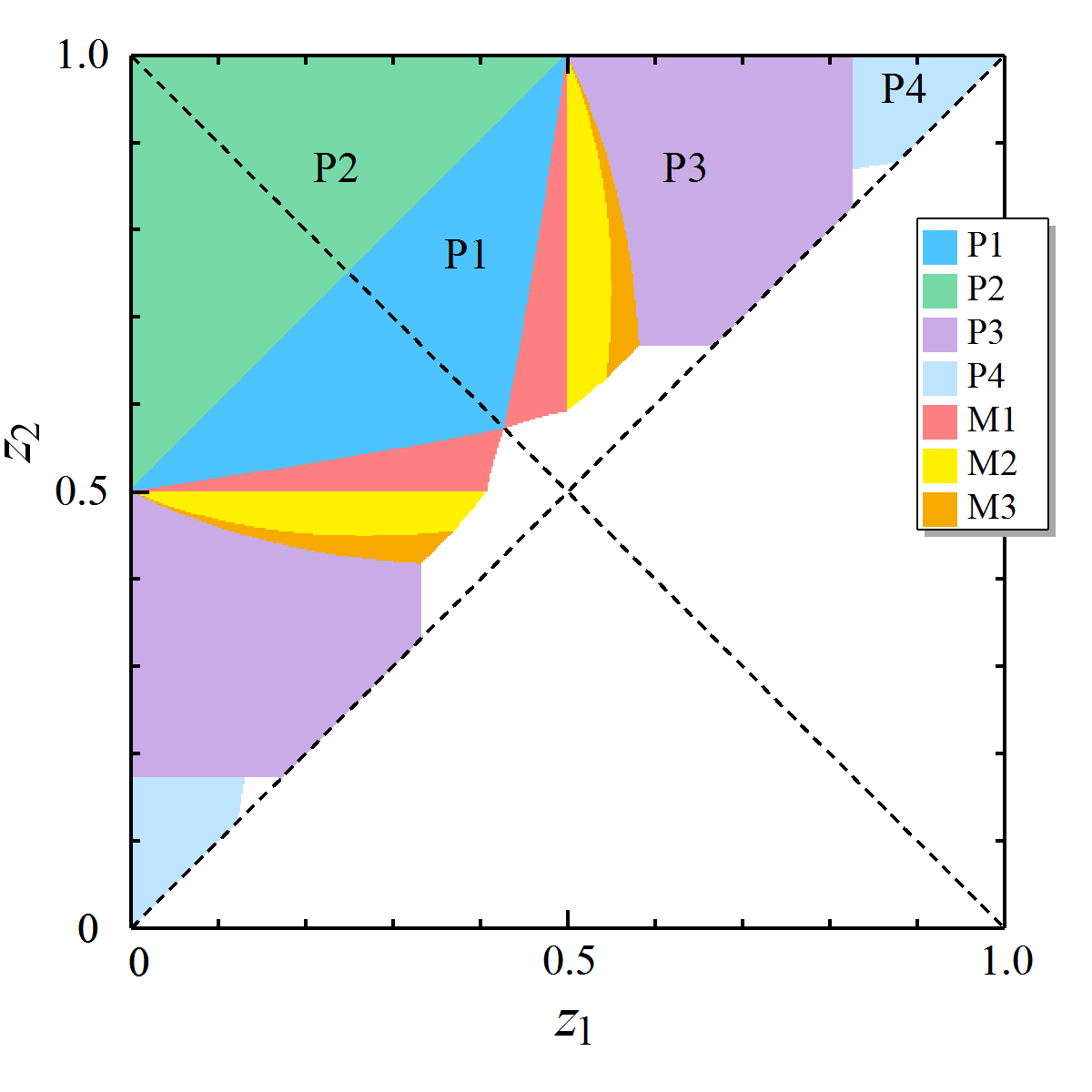}
\includegraphics[width=6cm]{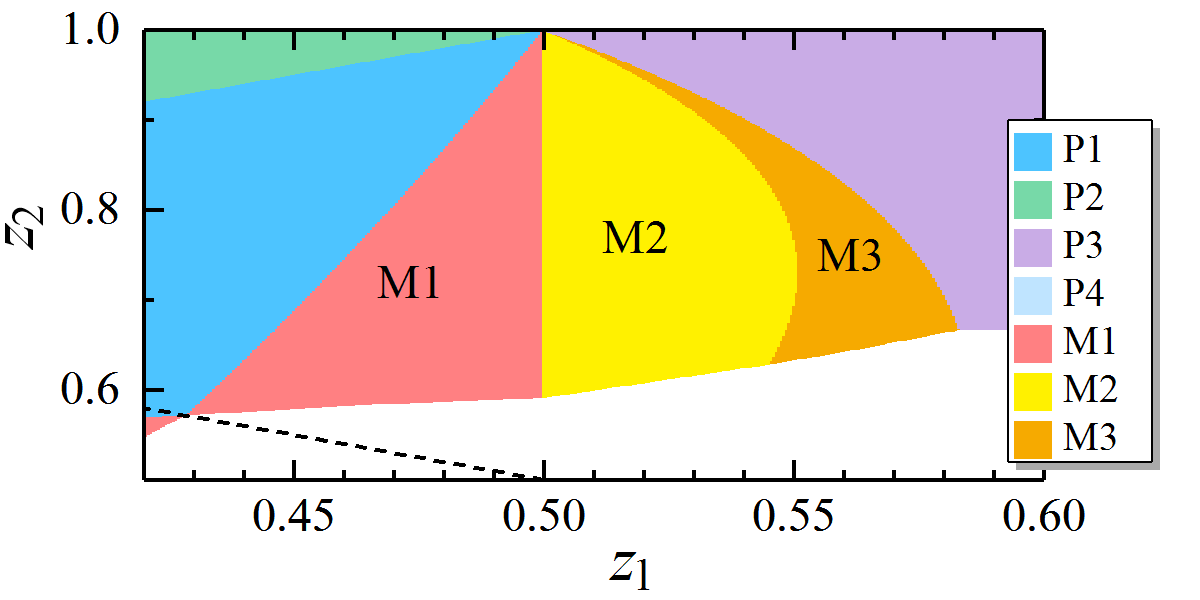}
\caption{Regions of different types of price equilibria in the $(z_1,z_2)$-plane.}
\label{fig:regions}
\end{figure}

\begin{figure}[ht]
\centering
\includegraphics[width=6cm]{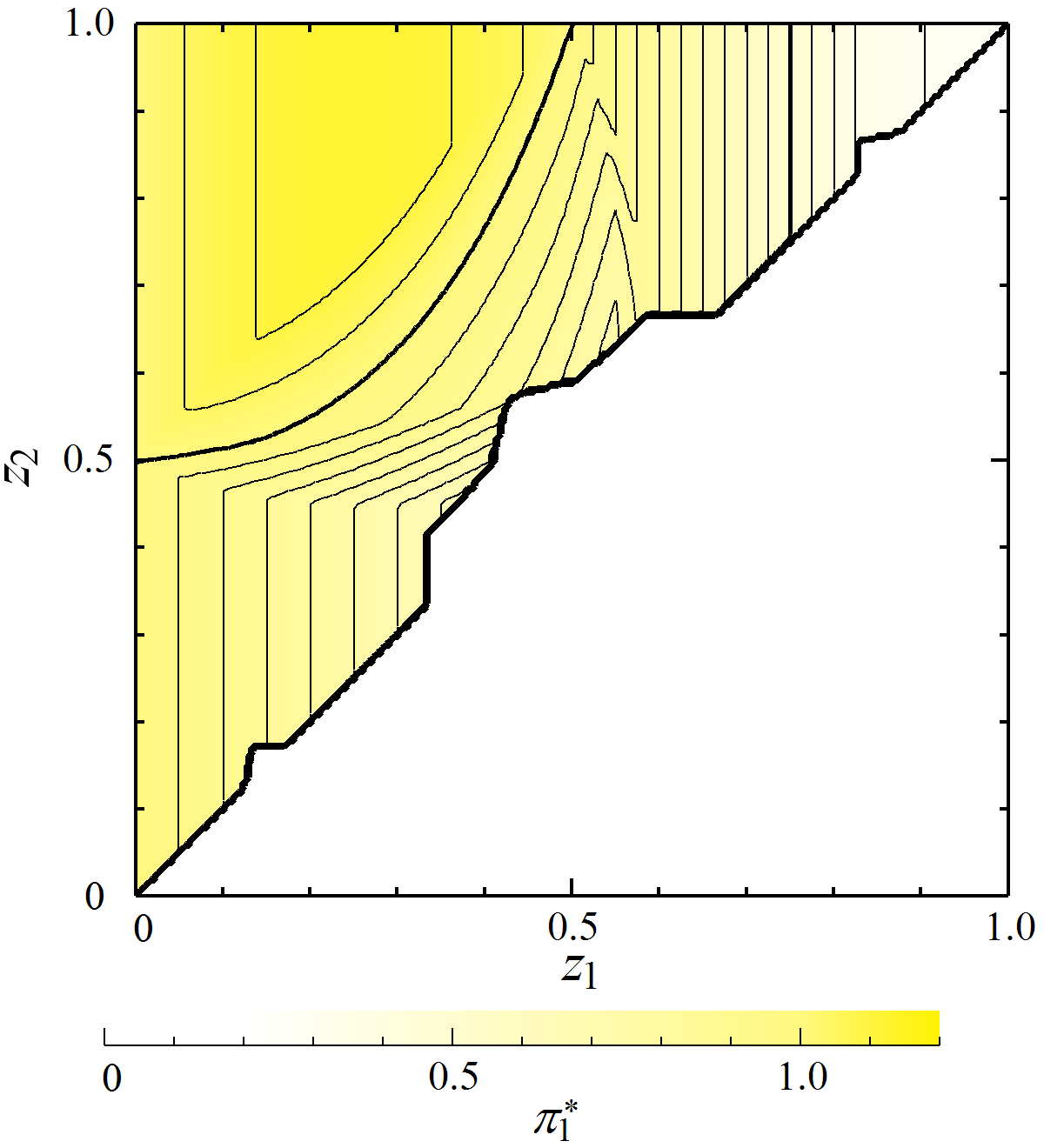}
\includegraphics[width=6cm]{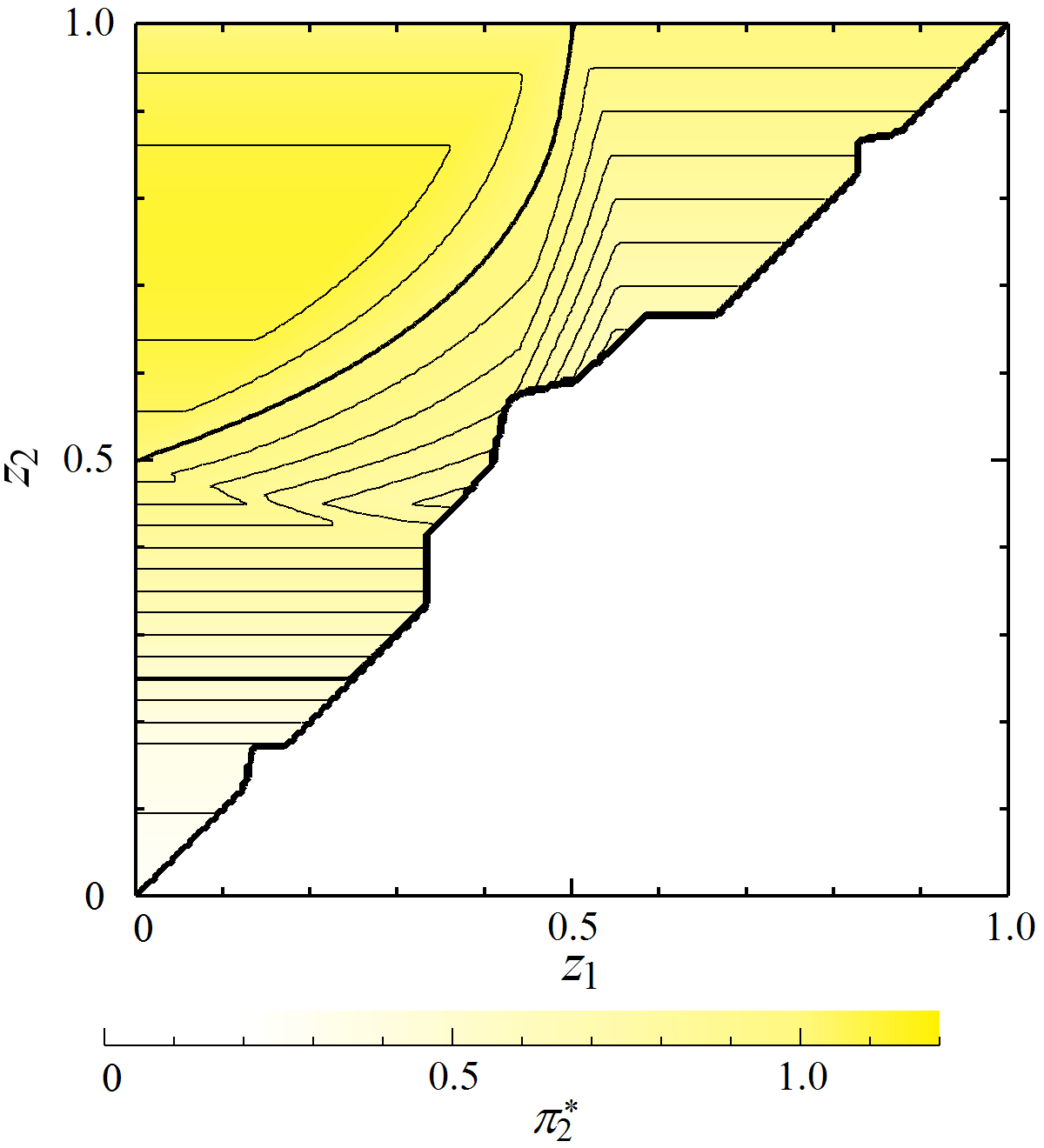}
\caption{Equilibrium profits $\pi_1^\star$ and $\pi_2^\star$ are plotted as functions of $(z_1,z_2)$.
The contours are spaced by 0.1. The thick contours are drawn at $\pi_j^\star = 0.5$ and $1$.
}
\label{fig:pi}
\end{figure}

We show the price equilibrium that corresponds to each region of the $(z_1,z_2)$ plane that is shown in Figure~\ref{fig:regions}. Figure~\ref{fig:pi} shows the expected profit obtained in the equilibrium. The present analysis focuses on the region $z_1<z_2$. Furthermore, since the problem has reflection symmetry with respect to the line $z_1+z_2=1$, the left-bottom region ($z_1<1/2$, $z_2<1/2$) is a mirror image of the right-top region ($z_1>1/2$, $z_2>1/2$). Note that below the $-45^{\circ}$ line, the positions of $z_1$ and $z_2$ are reversed in the figures below.

\section{Pure Strategy Equilibria}
This section focuses on a pure strategy equilibrium and shows some equilibria. In the beginning, we prepare for the finding of equilibria in this section. We obtain the following two lemmas. Hereinafter, Firm $ i$'s product will be referred to as $z_i$. See the Appendix~\ref{lemmaofcondition3appendix} for proof of these lemmas.
\begin{lemma}\label{prelemmaofcondition3}
If some informed consumers purchase the Firm $1$'s product, the others purchase the Firm $2$'s product, then $z_1 \ne z_2$ holds.
\end{lemma}
If $p_1+z_2-z_1=p_2+z_2-z_2$, the informed consumer at $t=z_2$ is indifferent between $z_1$ and $z_2$. In this case, no informed consumer except for $t=z_2$ purchases $z_2$. Then, we have the following equation about the informed consumer at the ideal point $t=z_2$;
\begin{equation}\label{condition3}
p_1-p_2+z_2-z_1=0.
\end{equation}
Now, we obtain the next lemma.
\begin{lemma}\label{lemmaofcondition3}
If \Eq{condition3} holds, every informed consumer at $t \ge z_2$ is indifferent between $z_1$ and $z_2$.
\end{lemma}
By Lemma~\ref{lemmaofcondition3}, the informed $C_3$ consumers located $t \ge z_2$ are indifferent between $z_1$ and $z_2$ if \Eq{condition3} holds. Now we assume the purchasing behavior of the informed consumers who are located at $t \ge z_2$ when they are indifferent between $z_1$ and $z_2$, as follows:
\begin{assumption}
The demand from those who are located on the interval $(t,1)$ is equally split between both firms. 
\end{assumption}

Here, we consider the case where all informed consumers purchase the Firm $1$'s product.
\begin{prop}\label{allshopperz1}
All of the informed consumers purchase the Firm $1$'s product when $p_1-p_2+z_2-z_1<0$ holds.
\end{prop}
Proof in Appendix~\ref{allshopperz1appendix}. Proposition~\ref{allshopperz1} details the condition for Firm $1$ to monopolize the informed market, and a similar condition exists for Firm $2$.

From now on, we will consider the case in which some informed consumers purchase the Firm $1$'s product while others purchase the Firm $2$'s product. 

We show a pure strategy equilibrium where both firms attract a segment of informed consumers over the interval $(0,1)$, and then they charge an equilibrium price pair $(p_1^{\star},p_2^{\star})=(1-z_1,z_2)$, respectively. 

In this equilibrium, both firm's profits are $\pi_1(1-z_1,z_2)=(1-z_1)(1+t)=(1-z_1)(\frac{1}{2}+z_1+z_2)$, $\pi_2(1-z_1,z_2)=(z_2)(1+(1-t))=(z_2)(\frac{5}{2}-z_1-z_2)$, respectively. 

We derive the condition that holds in this equilibrium heuristically and then obtain Proposition~\ref{case1characterized1}.

Intuitively, in this equilibrium, both firms have no incentive to attract all the informed consumers. The reason is that both firms are located far from $1/2$. In other words, they are located relatively close to their captive buyers. Thus, they can not take all informed consumers. This equilibrium is also similar to the pure strategy equilibrium found by \citeauthor{da}. However, we also show that this equilibrium is one of several pure strategy equilibria.

To show this equilibrium, we have to check two conditions, as follows: (1) Neither firm improves its profit when it slightly reduces $(\epsilon>0)$ its price. (2) Neither firm improves their profit when, in securing at least their captive buyers, $C_i$, they sharply discount their price to get all of the informed consumers, $C_3$, over the interval $(0,1)$.\footnote{Discounting sharply means that either firm captures all informed consumers. This second condition was related to a traditional topic found by \citeauthor{da}.} 

Here, we focus on the Firm $1$'s product case. The same argument applies to the Firm $2$'s case. Recall that there is a captive buyer in this model; the demand from captive buyers always guarantees a minimum profit, $\pi_i = 1$, for both firms at given location points. Both firms can obtain this profit by charging their prices $p_1=1-z_1$ and $p_2=z_2$, respectively, because $p_1+z_1=1$ and $p_2+(1-z_2)=1$.

First, we check condition (1). Suppose that $p_2=z_2$, we calculate Firm $1$'s profit when they charge $p_1 = p_1^{\star}-\epsilon$. We have $\pi_1(1-z_1-\epsilon,z_2)=(1-z_1)(\frac{1}{2}+z_1+z_2)-\epsilon(\frac{1}{2}+z_1+z_2-\frac{1-z_1}{2})-\frac{\epsilon^2}{2}$. We find that Firm $1$'s profit is decreased. Similarly, this argument is the same for firm $2$, and we have $\pi_2(1-z_1,z_2-\epsilon)=z_2(z_2+z_1-\frac{1}{2})-\epsilon(2-z_1-\frac{3}{2}z_2+\frac{1}{2})-\frac{\epsilon^2}{2}$. Thus, we show that condition (1) holds if they charge an equilibrium price pair $(p_1^{\star},p_2^{\star})=(1-z_1,z_2)$.

Next, we check condition (2). Note that, by Proposition~\ref{allshopperz1}, we have that $p_1 - p_2 + z_2 - z_1 < 0$ when an informed consumer $t=z_2$ purchases $z_1$. Suppose again that $p_2=z_2$, we substitute $p_2=z_2$ for $p_1 - p_2 + z_2 - z_1 < 0$, we have $p_1 < z_1$. Thus, if Firm $1$ discounts its price sharply, the supremum of its profit will be $\pi_1=z_1(1+1)=2z_1$.

At last, we check the condition that an informed consumer $t=z_1$ purchases $z_1$ in this equilibrium, in which either firm attracts a segment of the informed consumers. By $1-z_1+|z_1-z_1|<z_2+|z_1-z_2|$, we have that $\frac{1}{2}<z_2$. Similarly, we also check the condition that an informed consumer $t=z_2$ purchases $z_2$. We have that $\frac{1}{2}>z_1$. Therefore, $z_1<\frac{1}{2}$ and $\frac{1}{2}<z_2$ are necessary when both firms both firms split the informed market by charging an equilibrium price pair $(p_1^{\star},p_2^{\star})=(1-z_1,z_2)$.

We obtain the following Proposition~\ref{case1characterized1}. 
\begin{prop}[Pure equilibrium $1$, P1]\label{case1characterized1}
When $(z_1,z_2)$ satisfies the following equations
\begin{align}
&z_2 \le z_1+ \frac{1}{2}, \label{psubcond2-1}\\
&(1-z_1)(\frac{1}{2}+z_1+z_2) \ge 2z_1,\label{psubcond2-2}\\
&z_2(\frac{5}{2}-z_1-z_2) \ge 2(1-z_2),\label{psubcond2-3}
\end{align}
there exsits a price equilibrium such that $(p_1^{\star},p_2^{\star})=(1-z_1,z_2)$. Equilibrium profits are given by $\pi_1^\star = (1-z_1)(z_1+z_2+1/2)$ and $\pi_2^\star = z_2(5/2-z_1-z_2)$.
\end{prop}
Proof in Appendix~\ref{case1characterized1appendix}. Fig.~\ref{fig:priceP1} provides a schematic representation of Pure equilibrium $1$, P1.

The following equilibrium, characterized in Proposition~\ref{case1characterized2}, arises from the finite reservation utility. This equilibrium is similar to the pure strategy equilibrium found by \citeauthor{ECONOMIDES1984345}.
\begin{prop}[Pure equilibrium $2$, P2]\label{case1characterized2}
When $(z_1,z_2)$ satisfies the following equations
\begin{equation}
z_2 \ge z_1 + \frac{1}{2}
\end{equation}
there exsits a price equilibrium such that $(p_1^{\star},p_2^{\star})=(1-z_1,z_2)$. Equilibrium profits are given by $\pi_1^\star = (1-z_1)(1+2z_1)$ and $\pi_2^\star = z_2(3-2z_2)$.
\end{prop}
Proof in Appendix~\ref{case1characterized2appendix}. Fig.~\ref{fig:priceP2} provides a schematic representation of Pure equilibrium $2$, P2.

\begin{figure}[ht]
\centering
\includegraphics[width=6cm]{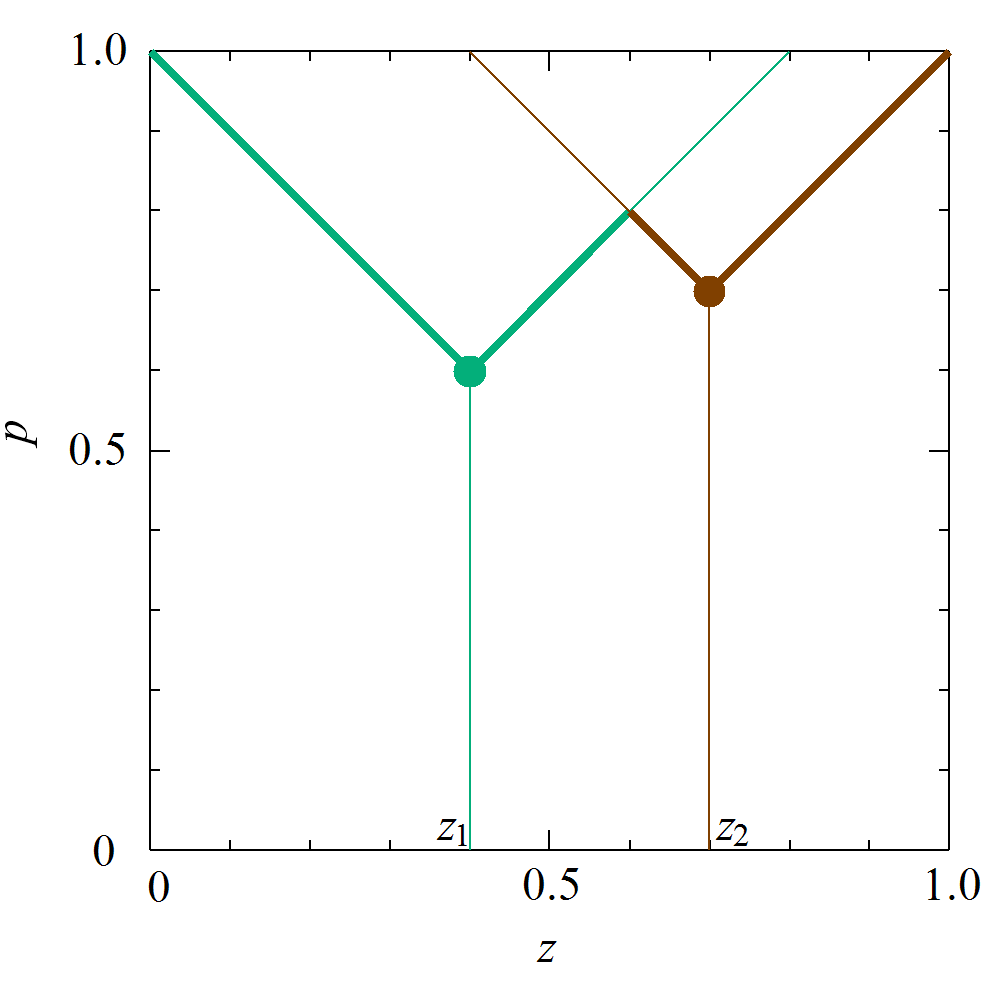}
\caption{Schematic for the pure strategy equilibrium P1, drawn for $(z_1,z_2)=(0.4, 0.7)$.}
\label{fig:priceP1}
\end{figure}      

\begin{figure}[htb]
\centering
\includegraphics[width=6cm]{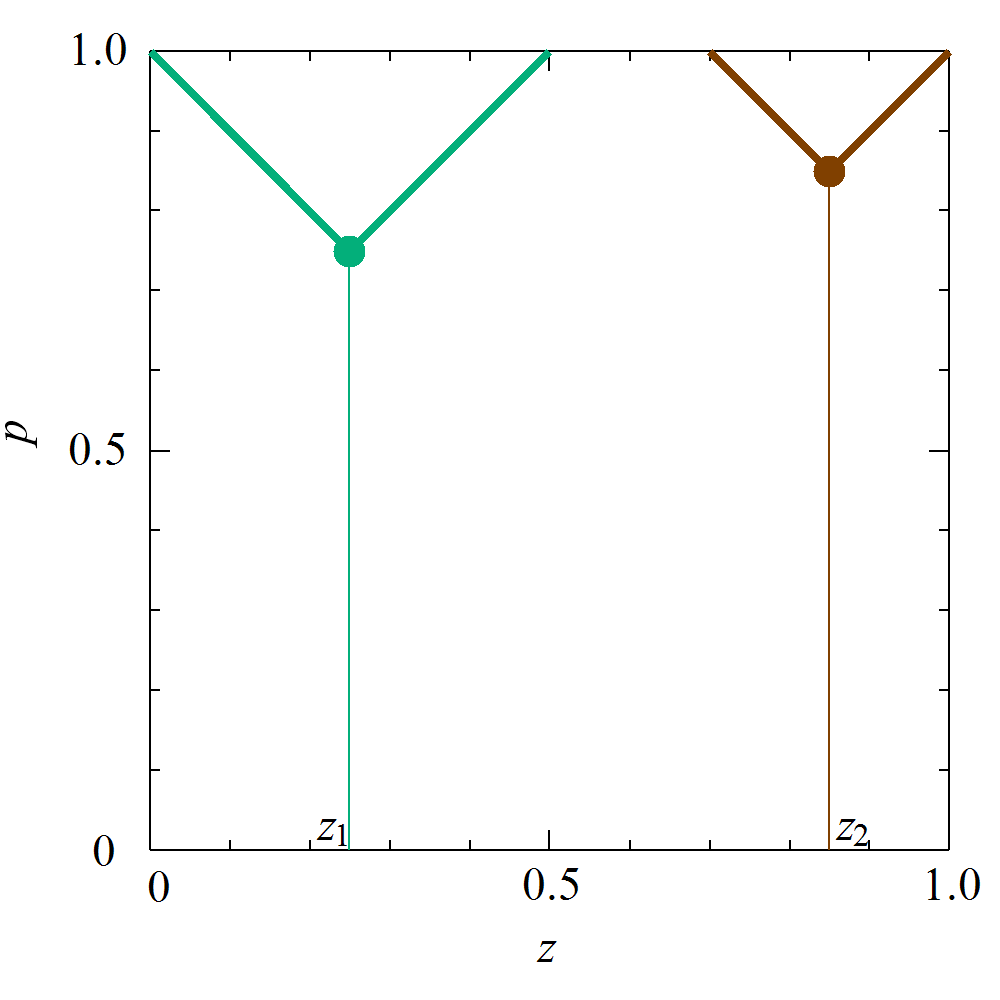}
\caption{Schematic for the pure strategy equilibrium P2, drawn for $(z_1,z_2)=(0.25, 0.85)$.}
\label{fig:priceP2}
\end{figure}

In the following, we show another pure strategy equilibrium for some $(z_1, z_2)$ pairs that do not satisfy \Eq{psubcond2-1}-\Eq{psubcond2-2}. We show two equilibria. The first equilibrium is an equilibrium where, for a given $(z_1,z_2), z_1 < z_2$, Firm $1$'s price $p_1=1-z_1$ and Firm $2$ do not compete in the informed market by $p_2=z_2$. In other words, Firm $1$ captures the entire informed market with its own uninformed reservation price. We obtain the following Proposition~\ref{pure4-1}.
\begin{prop}[Pure equilibrium $3$, P3]\label{pure4-1}
When $(z_1,z_2)$ satisfy the following equations 
\begin{align}
&(1-2z_1+z_2)(2-z_2) \le z_2, \label{pure1condi}\\
&z_1 \le 2(\sqrt{2}-1),\\
&2/3 \le z_2,
\end{align}
there exsits a pure strategy equilibrium such that $(p_1^{\star}, p_2^{\star})=(1-z_1, z_2)$. Equilibrium profits of both firms are $\pi_1^{\star} = 2(1-z_1), \pi_2^{\star}=z_2$ respectively.
\end{prop}
Proof in Appendix~\ref{pure4-1appendix}. Fig.~\ref{fig:priceP3} provides a schematic representation of Pure equilibrium $3$, P3.

\begin{figure}[htb]
\centering
\includegraphics[width=6cm]{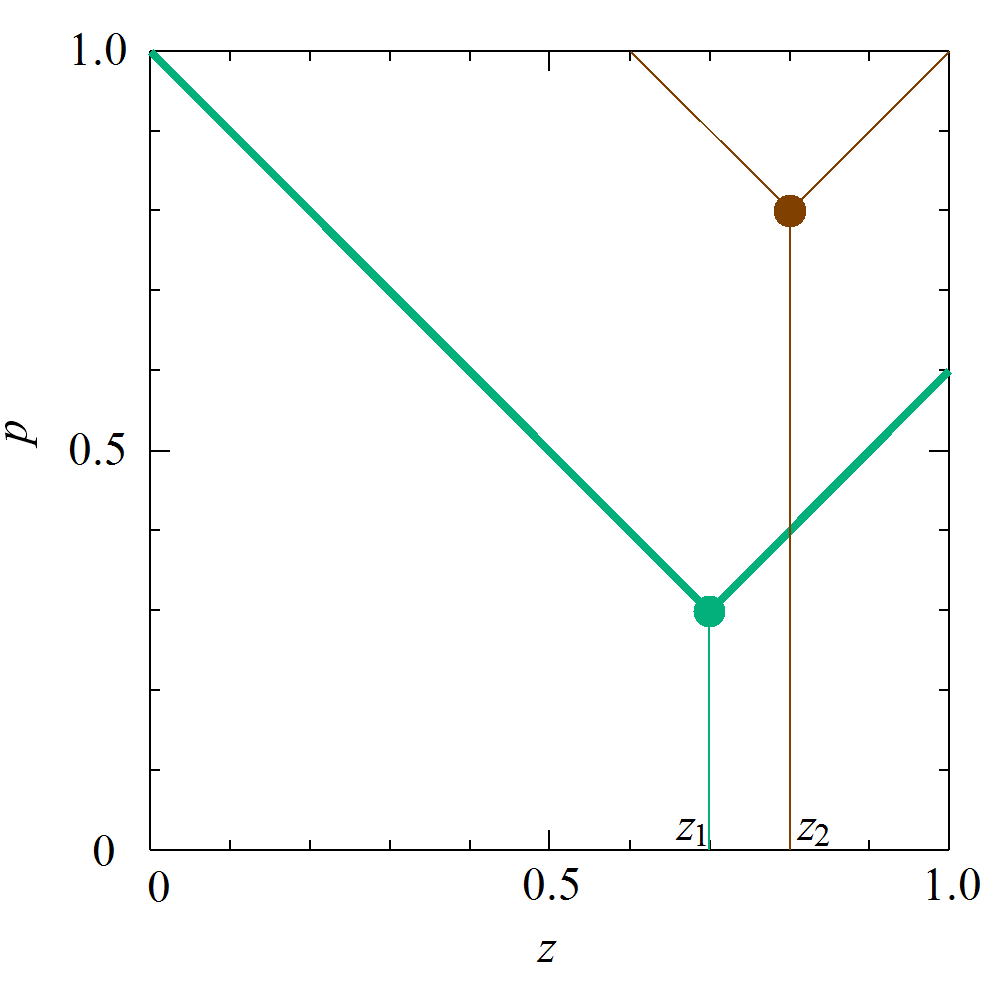}
\caption{Schematic for the pure strategy equilibrium P3, drawn for $(z_1,z_2)=(0.7, 0.8)$.}
\label{fig:priceP3}
\end{figure}      

The second equilibrium is also an equilibrium where for a given $(z_1,z_2), z_1 < z_2$, firm $2$ does not compete in the informed market with $p_2=z_2$. Thus, $(1-2z_1+z_2)(2-z_2) \le z_2$ and $1/2 \le z_1$. On the other hand, Firm $1$ only partially earns informed consumers on the left side of Firm $1$, $(0,z_1]$, because its equilibrium price exceeds its own uninformed reservation price. The condition on the marginal consumer $t \in (0,z_1]$ is $p_1+(z_1-t) \le 1$. Solve this and obtain $t \ge p_1-(1-z_1)$. Under this condition, we obtain the following Proposition~\ref{pure4-2}.

\begin{prop}[Pure equilibrium $4$, P4]\label{pure4-2}
When $(z_1,z_2)$ satisfy the following equations 
\begin{align}
% z_1 < z_2,\\
\left(1+\frac{z_1}{2}-z_2\right)\left(2-\frac{z_1}{2}\right) \le z_2,\label{pure2condi}\\
z_1 > 2\left(\sqrt{2}-1\right),
\end{align}
there exsits a pure strategy equilibrium such that $(p_1^{\star},p_2^{\star})=(1-\frac{z_1}{2}, z_2)$. Equilibrium profits of both firms are $\pi_1^{\star} = (1-\frac{z_1}{2})^2, \pi_2^{\star}=z_2$ respectively.
\end{prop}
Proof in Appendix~\ref{pure4-2appendix}. Fig.~\ref{fig:priceP4} provides a schematic representation of Pure equilibrium $4$, P4.

\begin{figure}[htb]
\centering
\includegraphics[width=6cm]{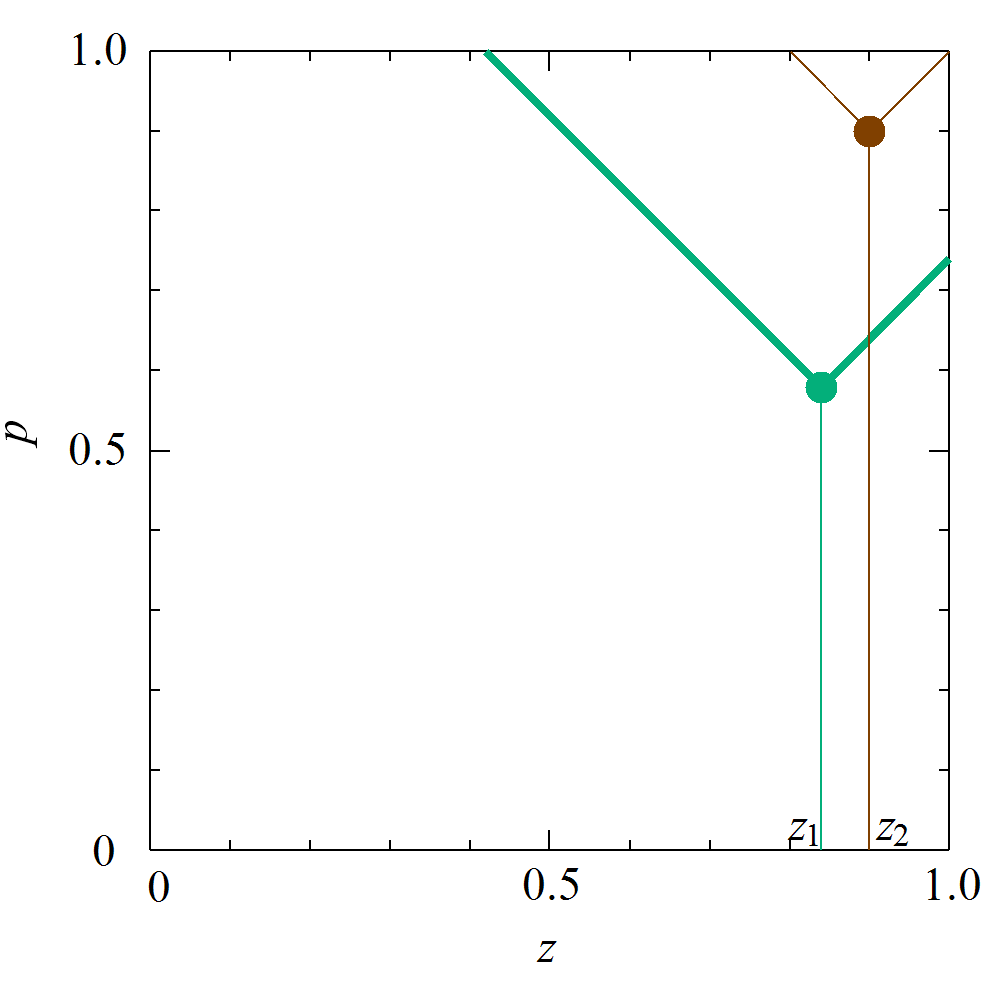}
\caption{Schematic for the pure strategy equilibrium P4, drawn for $(z_1,z_2)=(0.84, 0.9)$.}
\label{fig:priceP4}
\end{figure}      

\section{Mixed Strategy Equilibria}

% % 線形交通費用ゲーム（設定の説明は略）の「case 2」を解いている。その後の考察で，「case 2」は6個のcaseに分割されることが分かった。以下，順に解いていく。これらの均衡解は思ったよりも複雑で，方程式を演繹的に解いていくのは難しい。以下では，まず「天からのお告げ」によって解を与え，それが均衡戦略になっていることを証明するという手順をとる。
% \begin{comment}
% First, we define the equilibrium strategy. 
% \begin{dfn}
% Let $F_i^{\star}$, the cumulative probability distribution of prices $p_i$ of Firm $i$ ($i=1,2$), be an equilibrium strategy if it satisfies the following conditions
% \begin{align}
% &\mathrm{E}[\pi_1](p_1) = \pi_1^{\star} &(\mbox{for }p_1\in S_1),\\
% &\mathrm{E}[\pi_1](p_1) \leq \pi_1^{\star} &(\mbox{for }p_1\not\in S_1), \\
% &\mathrm{E}[\pi_2](p_2) = \pi_2^{\star} &(\mbox{for }p_2\in S_2),\\
% &\mathrm{E}[\pi_2](p_2) \leq \pi_2^{\star} &(\mbox{for }p_2\not\in S_2).
% \end{align}
% \end{dfn}
% Here, $\mathrm{E}[\pi_1](p_1)$ is "the expected profits of Firm $2$'s product when firm $1$ chooses $p_1$," which is defined as
% \begin{align}
% \mathrm{E}[\pi_1](p_1) = \int \pi_1(p_1,p_2) \thinspace \diff F_2(p_2)
% \end{align}
% The $\mathrm{E}[\pi_2](p_2)$ is defined similarly. Also, $\pi_1^{\star}$ and $\pi_2^{\star}$ denote the equilibrium profits. The sets $S_1$ and $S_2$ denote the supports of the price distributions of firms $1$ and $2$, respectively (the set of prices whose probability density is not zero).
% \end{comment}

In finding the equilibrium strategy below, the following function frequently appears in the solution. Therefore, we define it here.
\begin{dfn}[Exponential integral]
\begin{align}
\mathrm{Ei}(z) =-\int_{-z}^{\infty} \frac{\exp(-t)}{t} \diff t 
= \int_{-\infty}^{z} \frac{\exp(t)}{t} \diff t \label{eq:defEi}
\end{align}
\end{dfn}
% \begin{comment}
% \begin{lemma}\label{lemmaforeq:defEiareequal}
% The two integrals in \Eq{eq:defEi} are equal.
% \end{lemma}
% \begin{proof}
% In the first integral of \Eq{eq:defEi}, if $t=-u$, then
% \begin{align}
% -\int_{-z}^{\infty} \frac{\exp(-t)}{t} \diff t
% = -\int_{z}^{-\infty} \frac{\exp(u)}{(-u)} (-1)\diff u
% = \int_{-\infty}^{z} \frac{\exp(u)}{(-u)} (-1)\diff u
% = \int_{-\infty}^{z} \frac{\exp(u)}{u} \diff u
% \label{eq:prfEi}
% \end{align}
% Since any symbol can be used for the integral variable, rewriting the last integral in \Eq{eq:prfEi}, we have
% \begin{align}
% \int_{-\infty}^{z} \frac{\exp(t)}{t} \diff t
% \end{align}
% \end{proof}
% \end{comment}
The following Lemma is trivial from the definition of $\mathrm{Ei}(z)$.
\begin{lemma}
\begin{align}
\frac{\diff}{\diff z} \mathrm{Ei}(z) = \frac{\exp(z)}{z}
\label{eq:diffEi}
\end{align}
\end{lemma}
Hereinafter, we use the following symbol, $\lambda_2$, to simplify the notation.
\begin{dfn}
\begin{align}
& \lmb \defeq \frac{1}{2(1-z_2)} \label{eq:lmbdef}\\
& \de \defeq z_2 - z_1 \label{eq:dedef}
\end{align}
\end{dfn}

\begin{dfn}
\begin{align}
& g_0(p ; b)  \defeq 1 + \frac{2}{ (3 - z_1 - 2 z_2 - b) } \exp\left(-\lmb(b-p)\right) \\
& \begin{aligned} 
    g_\pi(p; b) \defeq &
    \; { \frac{2\lmb}{ p+\de} } 
    - \frac{2\lmb}{ (b+\de) } \frac{(1-z_1-b)}{ (3 - z_1 - 2 z_2 - b) } \exp\left(-\lmb(b-p)\right) \cr
    & {} - 2 \lmb^2 \exp\left(\lmb(p+\de)\right) 
    \Bigl( \mathrm{Ei}\left( -\lmb (b + \de ) \right) - \mathrm{Ei}\left( -\lmb(p+\de)\right) \Bigr) 
    \end{aligned} \\
& g(p ; b, \pi)  \defeq  g_0(p; b) - \pi \, g_\pi(p; b) \\
& \begin{aligned}
    h(p ; a)  \defeq &
    \;   4 \lmb \exp\left( -\lmb(p-a)\right)  - \frac{ 4\lmb (a-\de) }{ (p-\de) } \cr
    & {} + 4 \lmb^2 (a-\de) \exp\left(-\lmb(p-\de)\right) 
    \Bigl( \mathrm{Ei}\left( \lmb (p - \de ) \right) - \mathrm{Ei}\left( \lmb(a-\de)\right) \Bigr)
    \end{aligned} 
\end{align}
\end{dfn}
\noindent
Here, the symbol $g(p; b,\pi)$  is defined as a function of $p$ that contains $b$ and $\pi$ as parameters.
Similarly, the symbol $h(p; a)$ is a function of $p$ that contains $a$ as a parameter.
Then, straightforward calculation shows the following properties.
\begin{lemma}
\begin{align}
& g(p; b, \pi) - \frac{1}{\lmb} \frac{\diff}{\diff p} g(p; b, \pi)  = 1 - \frac{ 2 \pi } { (p+\de)^2 } \label{eq:gdiff} \\
& g(b; b, \pi ) = \frac{ 1 }{(3-z_1-2z_2-b)}
\left( \left( 5-z_1-2z_2-b \right)
- \frac{2\pi}{(b+\de)}
\right)
\label{eq:gbb} \\
& h(p; a) + \frac{1}{\lmb} \frac{\diff}{\diff p} h(p; a)  =  \frac{ 4(a-\de) } { (p-\de)^2 } \label{eq:hdiff}\\
& h(a; a) = 0 \label{eq:haa0}
\end{align}
\end{lemma}

% \begin{figure}[ht]
% \centering
% \includegraphics[width=6cm]{figures/map_w_blowup.png}
% \caption{The width $w$ of the equilibrium distributions is plotted as a function of $(z_1,z_2)$.
% The contours are spaced by 0.01. 
% }
% \label{fig:w}
% \end{figure}

\subsection{Mixed Strategy Equilibrium $1$}

Suppose the location pair $(z_1,z_2)$ is in the range satisfying the following.
\begin{align}
& z_1 \le 1/2 \label{eq:rgn1a} \\
& z_1+z_2 \ge 1 \label{eq:rgn1b} \\
& \left( 1-z_1 \right) \left( \frac{1}{2} + z_1 + z_2 \right) < 2z_1 \label{eq:rgn1c}
\end{align}
Suppose the following equation for $w$ has a positive solution $w>0$.
\begin{align}
\frac{1}{ 2 \lmb } h(z_2; z_2-w) + z_1 + z_2 + \frac{1}{2} - \frac{ 2w }{ z_1 } = \frac{ 2 (z_1-w) }{ (1-z_1) }
\label{eq:defw1}
\end{align}
We further require that the location pair $(z_1,z_2)$ satisfies the following inequality. 
\begin{align}
& 2 - \frac{w}{2} - \left( \frac{3}{2} - z_2 \right) g_0(z_1; z_1) \cr
&{} + \left( \frac{w}{z_2(z_2-w)} + \left( \frac{3}{2} - z_2 \right) g_\pi (z_1; z_1) - \frac{1}{(1-z_2)} \right) 
\frac{ g_0(z_1-w;z_1) }{ g_\pi(z_1-w;z_1) } \leq 0
\label{eq:condw1}
\end{align}

\begin{prop}[Mixed Strategy Equilibrium $1$]\label{thm:sol1}

Suppose that the location pair $(z_1,z_2)$ is inside of the region given by \Eq{eq:rgn1a}--(\ref{eq:rgn1c}), and that \Eq{eq:defw1} has a solution $w$ that satisfies the inequality \Eq{eq:condw1}
Then, the mixed strategy given by the following cumulative distribution function is an equilibrium,
\begin{align}
&F_1^\star(p_1) =
\begin{dcases*}
0 & (for $p_1\le z_1-w$) \\
g(p_1; z_1, \pi_2^\star)
& (for $z_1-w \le p_1 \le z_1$ ) \\
g(z_1; z_1, \pi_2^\star) & (for $z_1 \le p_1 < 1-z_1$ ) \\
1 & (for $p_1 > 1-z_1$ )
\end{dcases*} \label{eq:F1sol1}\\
\vspace{2\baselineskip}\cr
&F_2^\star(p_2) =
\begin{dcases*}
0 & (for $p_2\le z_2-w$) \\
h(p_2; z_2-w) 
& (for $z_2-w \le p_2 < z_2$ ) \\
1 & (for $p_2 > z_2$ )
\end{dcases*} \label{eq:F2sol1}
\end{align}
with the equilibrium profits given by the following.
\begin{align}
& \pi_1^\star = 2(z_1-w) 
\label{eq:pi1e1} \\
& \pi_2^\star = \frac{ g_0(z_1-w;z_1) }{ g_\pi(z_1-w;z_1) } 
\label{eq:pi2e1}
\end{align}
\end{prop}

The supports of these distributions are as follows.
\begin{align}
& \mathrm{supp} \  F_1^\star = [z_1-w, z_1] \cup \{ 1-z_1 \} \label{eq:supp1sol1} \\
& \mathrm{supp} \  F_2^\star = [z_2-w, z_2] \label{eq:supp2sol1}
\end{align}
The distribution $F_1$ of Firm $1$ is continuous at $p_1 = z_1-w$, as ensured by \Eq{eq:pi2e1}, and at $p_1=z_1$ as trivially seen from \Eq{eq:F1sol1}. It has an atom at $p_1=1-z_1$. The distribution $F_2$ of Firm $2$ is continuous at $p_2=z_2-w$ due to \Eq{eq:haa0}. It has an atom at $p_2=z_2$.

The expected profit $\mathrm{E}[\pi_i](p_i)$ of Firm $i$ when it chooses a price $p_i$ is given by the following integration over the price distribution of its counterpart.
\begin{align}
& \mathrm{E}[\pi_1](p_1) = p_1 \left( 1 + \int_{p_1-\de}^{p_1+\de} \frac{ z_1+z_2-p_1+p_2}{2} \diff F_2^\star(p_2) + \left( 1 - F_2^\star(p_1+\de) \right)  \right) \label{eq:expect1} \\
& \mathrm{E}[\pi_2](p_2) = p_2 \left( 1 + \int_{p_2-\de}^{p_2+\de} \frac{ z_1+z_2+p_1-p_2}{2} \diff F_1^\star(p_1) + \left( 1 - F_1^\star(p_2+\de) \right)  \right) \label{eq:expect2}
\end{align}
That $\mathrm{E}[\pi_1](p_1) = \pi_1^\star$ for $z_1-w\leq p_1 \leq z_1$ can be proved by using \Eq{eq:hdiff}. That $\mathrm{E}[\pi_1](p_1) = \pi_1^\star$ for $p_1 = 1 - z_1$ is ensured owing to \Eq{eq:defw1}. That $\mathrm{E}[\pi_2](p_2) = \pi_2^\star$ for $z_2-w\leq p_2 \leq z_2$ can be proved by using \Eq{eq:gdiff}. Finally, \Eq{eq:condw1} ensures that $\mathrm{E}[\pi_2](1-z_2) < \pi_2^\star$ so that Firm 2 has no incentive to choose the price $1-z_2$, which is outside the support given by \Eq{eq:supp2sol1}. This completes the proof that the probability distributions given by \Eq{eq:F1sol1} and \Eq{eq:F2sol1} are a price equilibrium. See Appendix~\ref{Mix1proof} for details of the proofs.

Note that for the Proposition~\ref{thm:sol1} to hold, not only must $(z_1,z_2)$ be in the range of equation \Eq{eq:rgn1a}--\Eq{eq:rgn1c}, but the condition in equation \Eq{eq:condw1} must hold. The numerical solution of $(z_1,z_2)$ where Proposition~\ref{thm:sol1} holds is shown in Figure~\ref{fig:regions}. 

\begin{figure}[ht]
\centering
\includegraphics[width=6cm]{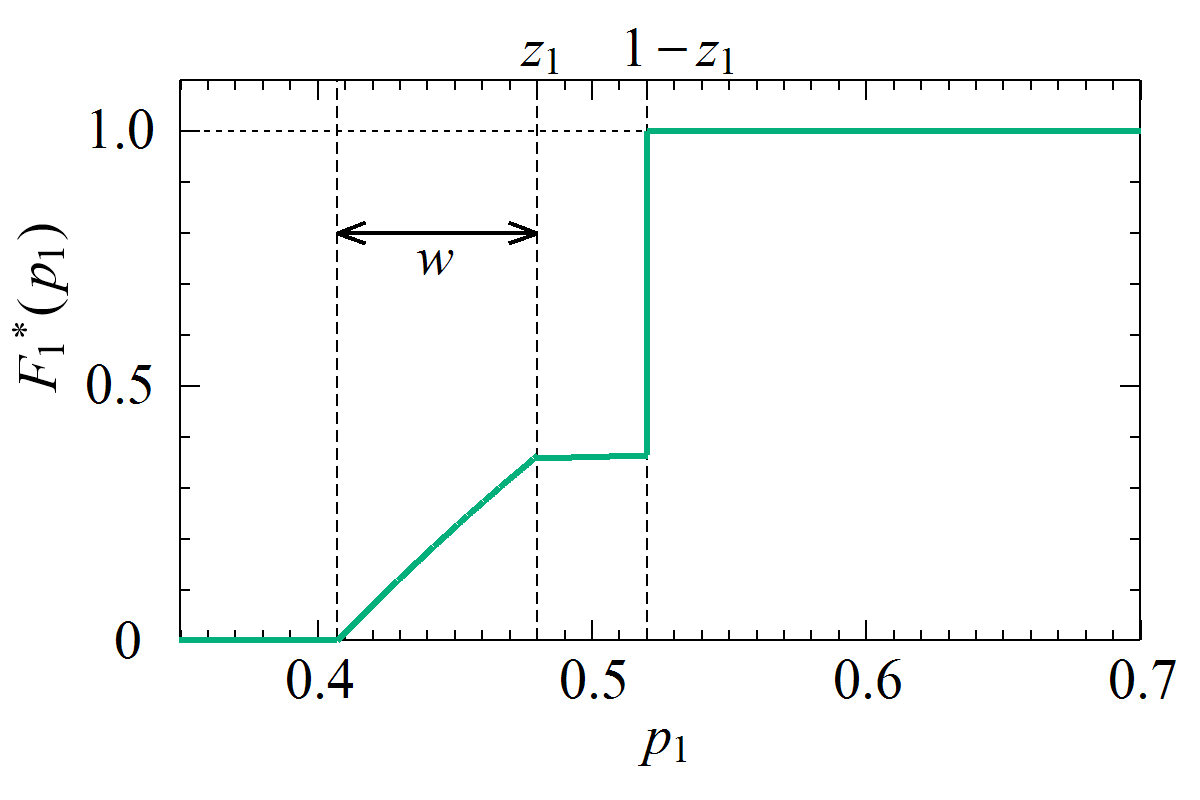}
\includegraphics[width=6cm]{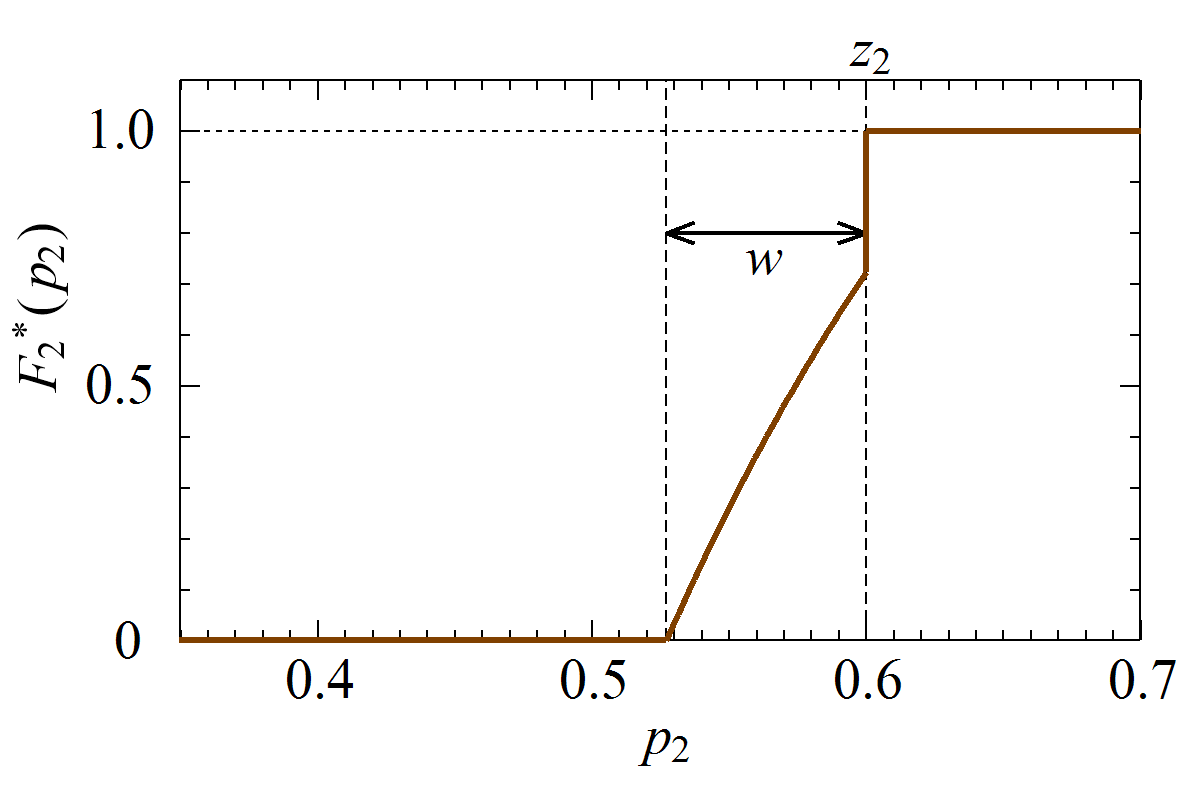}
\caption{Cumulative price distribution functions $F_1^\star$ (left) and $F_2^\star$ (right) for the mixed strategy equilibrium M1, drawn for $(z_1,z_2)=(0.48, 0.6)$.}
\label{fig:CDF_M1}
\end{figure}

\subsection{Mixed Strategy Equilibrium $2$}

Here, we suppose that the location pair $(z_1,z_2)$ is in the following range.
\begin{align}
& 1/2 \le z_1 < z_2 \label{eq:rgn2a} \\
& \left( 1-2z_1+z_2 \right) \left( 2 - z_2 \right) > z_2 \label{eq:rgn2b}
\end{align}
And suppose the following equation for $w$ has a positive solution $w>0$.
\begin{align}
h(1-2z_1+z_2; 1-2z_1+z_2-w) = 1
\label{eq:defw2}
\end{align}
We further restrict the range of $(z_1,z_2)$ so that the following two inequalities are satisfied. 
\begin{align}
& \left(z_1 - (1-z_2)(z_2-z_1)  + \frac{(1-z_2)w}{2} \right) \cr
& {} \ge  \left( \frac{z_1}{(1-2z_1+z_2)} + \frac{(1-z_2)}{(1-2z_1+z_2-w)} - 1\right)  
\frac{ g_0( 1-z_1-w; 1-z_1) }{ g_\pi( 1-z_1-w; 1-z_1) }
\label{eq:condw2a} \\
& \frac{ g_0( 1-z_1-w; 1-z_1) }{ g_\pi( 1-z_1-w; 1-z_1) } \geq z_2 
\label{eq:condw2b}
\end{align}

\begin{prop}[Mixed Strategy Equilibrium $2$]\label{thm:sol2}
Suppose that the location pair $(z_1,z_2)$ is inside of the region given by \Eq{eq:rgn2a}--(\ref{eq:rgn2b}), and that \Eq{eq:defw2} has a solution $w$ that satisfies the inequality \Eq{eq:condw2a} and \Eq{eq:condw2b}.
Then, the mixed strategy given by the following cumulative distribution function is an equilibrium,
\begin{align}
&F_1^\star(p_1) =
\begin{dcases*}
0 & (for $p_1\le 1-z_1-w$) \\
g(p_1; 1-z_1, \pi_2^\star)
& (for $1-z_1-w \le p_1 < 1-z_1$ ) \\
1 & (for $p_1 > 1-z_1$ )
\end{dcases*} \label{eq:F1sol2}\\
\vspace{2\baselineskip}\cr
&F_2^\star(p_2) =
\begin{dcases*}
0 & (for $p_2\le 1 - 2z_1 + z_2-w$) \\
h(p_2; 1-2z_1+z_2-w)
& (for $1-2z_1+z_2-w \le p_2 \le 1-2z_1+z_2$ ) \\
1 & (for $p_2 \ge 1-2z_1+z_2 $ )
\end{dcases*} \label{eq:F2sol2}
\end{align}
with the equilibrium profits given by the following.
\begin{align}
& \pi_1^\star = 2(1 - z_1-w) 
\label{eq:pi1e2} \\
& \pi_2^\star = \frac{ g_0( 1-z_1-w; 1-z_1) }{ g_\pi( 1-z_1-w; 1-z_1) }
\label{eq:pi2e2}
\end{align}
\end{prop}

The supports of $F_1^\star$ and $F_2^\star$ are $1-z_1-w\leq p_1 \leq 1-z_1$ and $1-2z_1+z_2-w \leq p_2 \leq 1-2z_1+z_2$, respectively.
The distribution $F_1^\star$ of Firm 1 is continuous at $p_1 = 1-z_1-w$, as ensured by \Eq{eq:pi2e2}. It has an atom at $p_1=1-z_1$. The distribution $F_2^\star$ of Firm 2 is continuous at $p_2=1-2z_1+z_2-w$ owing to \Eq{eq:haa0}, and at  $p_2=1-2z_1+z_2$ owing to \Eq{eq:defw2}. 

That $\mathrm{E}[\pi_1](p_1) = \pi_1^\star$ for $1-z_1-w\leq p_1 \leq 1-z_1$ can be proved by using \Eq{eq:hdiff}. That $\mathrm{E}[\pi_2](p_2) = \pi_2^\star$ for $z_2-w\leq p_2 \leq z_2$ can be proved by using \Eq{eq:gdiff}. \Eq{eq:condw2b} ensures that $\pi_2^\star \geq z_2$ due to \Eq{eq:pi2e2}, so that Firm 2 has no incentive to choose the price $p_2=z_2$, in which case it can gain only from its fan and the profit would be $z_2$. 
Finally, \Eq{eq:condw2a} ensures that $\mathrm{E}[\pi_2](1-z_2) < \pi_2^\star$ so that Firm 2 has no incentive to choose the price lower than $1-z_2$ to win the shoppers against Firm 1. This completes the proof that the probability distributions given by \Eq{eq:F1sol2} and \Eq{eq:F2sol2} are a price equilibrium.  See Appendix~\ref{Mix2proof} for details of the proofs.

Note that for the Proposition~\ref{thm:sol2} to hold, not only must $(z_1,z_2)$ be in the range of equation \Eq{eq:rgn2a}--\Eq{eq:rgn2b}, but the condition in both equation \Eq{eq:condw2a} and \Eq{eq:condw2b} must hold. The numerical solution of $(z_1,z_2)$ where Proposition~\ref{thm:sol2} holds is shown in Figure~\ref{fig:regions}. 
% in Appendix~\ref{mixfigures}.

\begin{figure}[ht]
\centering
\includegraphics[width=6cm]{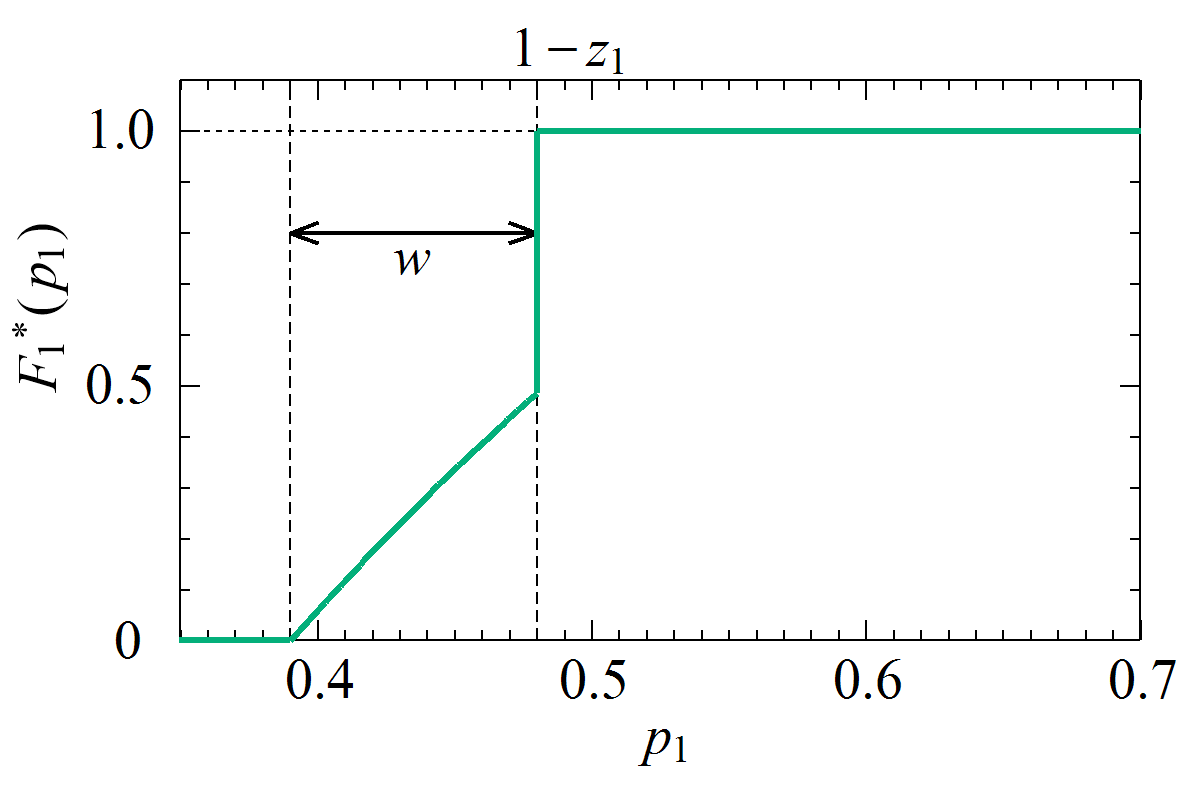}
\includegraphics[width=6cm]{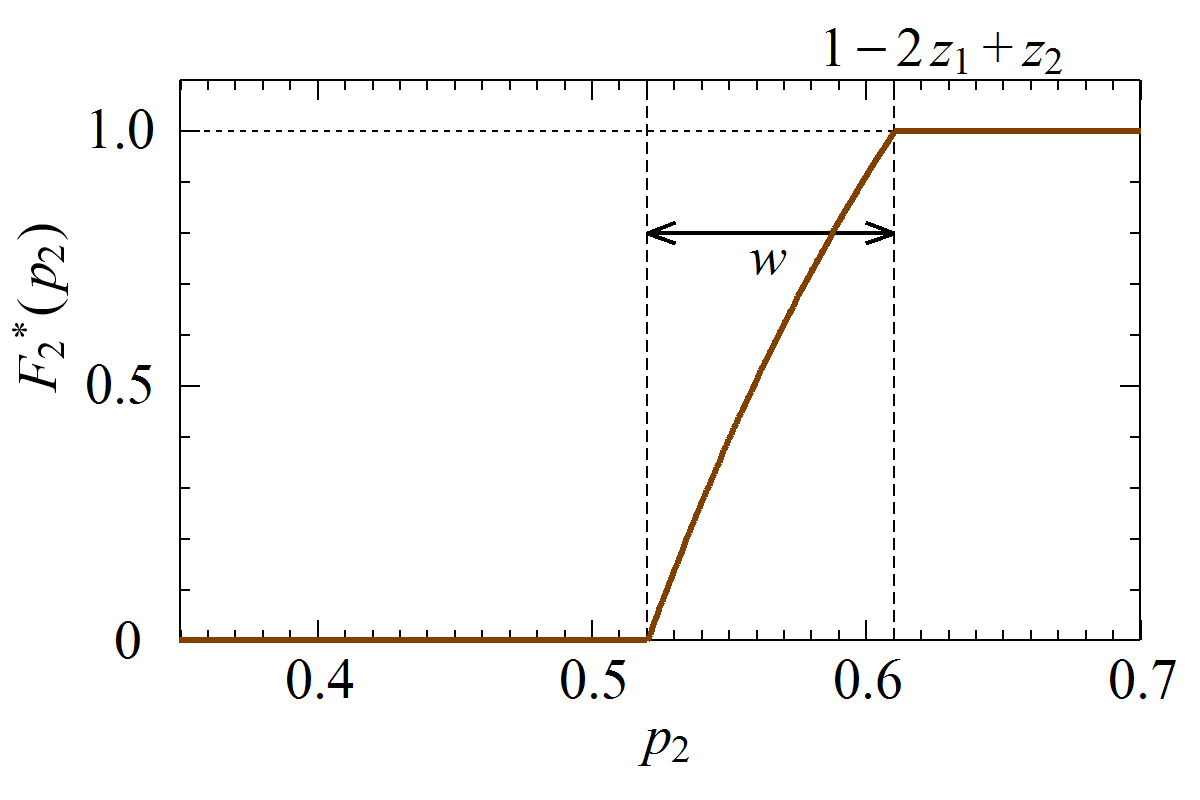}
\caption{Cumulative price distribution functions $F_1^\star$ (left) and $F_2^\star$ (right) for the mixed strategy equilibrium M2, drawn for $(z_1,z_2)=(0.52, 0.65)$.}
\label{fig:CDF_M2}
\end{figure}

\subsection{Mixed Strategy Equilibrium $3$}

Suppose the location pair $(z_1,z_2)$ is in the following range.
\begin{align}
& 1/2 \le z_1 < z_2 \label{eq:rgn3a} \\
& \left( 1-2z_1+z_2 \right) \left( 2 - z_2 \right) > z_2 \label{eq:rgn3b}
\end{align}
Suppose the following equation for $w$ has a positive solution $w>0$.
\begin{align}
g_0(1-z_1-w; 1-z_1) - z_2 \, g_\pi(1-z_1-w; 1-z_1) = 0
\label{eq:defw3}
\end{align}
We further restrict the range of $(z_1,z_2)$ so that the following two inequalities are satisfied. 
\begin{align}
& 
h(1-2z_1+z_2; 1-2z_1+z_2-w) \leq 1
\label{eq:condw3a} \\
& 
(z_2-z_1)z_2 + 2z_1
+ \frac{(1-z_2)w}{2} - \frac{z_1z_2}{( 1-2z_1+z_2 )} - \frac{(1-z_2)z_2}{  (1-2z_1+z_2-w) }
\geq 0
\label{eq:condw3b}  
\end{align}

\begin{prop}[Mixed Strategy Equilibrium $3$]\label{thm:sol3}

Suppose that the location pair $(z_1,z_2)$ is inside of the region given by \Eq{eq:rgn3a}--(\ref{eq:rgn3b}), and that \Eq{eq:defw3} has a solution $w$ that satisfies the inequality \Eq{eq:condw3a} and \Eq{eq:condw3b}.
Then, the mixed strategy given by the following cumulative distribution function is an equilibrium,
\begin{align}
&F_1^\star(p_1) =
\begin{dcases*}
0 & (for $p_1\le 1-z_1-w$) \\
g(p_1; 1-z_1, \pi_2^\star)
& (for $1-z_1-w \le p_1 < 1-z_1$ ) \\
1 & (for $p_1 > 1-z_1$ )
\end{dcases*} \label{eq:F1sol3}\\
\vspace{2\baselineskip}\cr
&F_2^\star(p_2) =
\begin{dcases*}
0 & (for $p_2\le 1 - 2z_1 + z_2-w$) \\
h(p_2; 1-2z_1+z_2-w)
& (for $1-2z_1+z_2-w \le p_2 \le 1-2z_1+z_2$ ) \\
h(1-2z_1+z_2; 1-2z_1+z_2-w)
& (for $1-2z_1+z_2 \le p_2 < z_2  $ ) \\
1 & (for $p_2 > z_2 $ )
\end{dcases*} \label{eq:F2sol3}
\end{align}
with the equilibrium profits given by the following.
\begin{align}
& \pi_1^\star = 2(1-z_1-w) 
\label{eq:pi1e3} \\
& \pi_2^\star = z_2
\label{eq:pi2e3}
\end{align}
\end{prop}

The supports of these distributions are as follows.
\begin{align}
& \mathrm{supp} \  F_1^\star = [1-z_1-w, 1-z_1] \label{eq:supp1sol3} \\
& \mathrm{supp} \  F_2^\star = [1-2z_1+z_2-w, 1-2z_1+z_2]  \cup \{ z_2 \} \label{eq:supp2sol3}
\end{align}
The distribution $F_1^\star$ of Firm 1 is continuous at $p_1 = 1 - z_1-w$, as ensured by \Eq{eq:defw3}. It has an atom at $p_1=1-z_1$. The distribution $F_2^\star$ of Firm 2 is continuous at $p_2=1-2z_1+z_2-w$ due to \Eq{eq:haa0}, and at $p_2=1-2z_1+z_2$ as trivially seen from \Eq{eq:F2sol3}. It has an atom at $p_2=z_2$.

That $\mathrm{E}[\pi_1](p_1) = \pi_1^\star$ for $1-z_1-w\leq p_1 \leq 1-z_1$ can be proved by using \Eq{eq:hdiff}. That $\mathrm{E}[\pi_2](p_2) = \pi_2^\star$ for $1-2z_1+z_2-w\leq p_2 \leq 1-2z_1+z_2$ can be proved by using \Eq{eq:gdiff}. That $\mathrm{E}[\pi_2](p_2) = \pi_2^\star$ for $p_2 = z_2$ can be easily proved by noting that the whole support of $F_1$ is contained in $p_1<z_2-\de=z_1$ and therefore Firm 2 can gain only from its fan when $p_2=z_2$. 
Finally, \Eq{eq:condw3a} ensures that $\mathrm{E}[\pi_2](1-z_2) < \pi_2^\star$ so that Firm 2 has no incentive to choose the price lower than $1-z_2$ to win the shoppers against Firm 1. This completes the proof that the probability distributions given by \Eq{eq:F1sol3} and \Eq{eq:F2sol3} are a price equilibrium. See Appendix~\ref{Mix3proof} for details of the proofs.

Note that for the Proposition~\ref{thm:sol3} to hold, not only must $(z_1,z_2)$ be in the range of \Eq{eq:rgn3a}--\Eq{eq:rgn3b}, but the condition in \Eq{eq:condw3a} and \Eq{eq:condw3b} must hold. The numerical solution of $(z_1,z_2)$ where Proposition~\ref{thm:sol3} holds is shown in Figure~\ref{fig:regions}.
% in Appendix~\ref{mixfigures}.

\begin{figure}[ht]
\centering
\includegraphics[width=6cm]{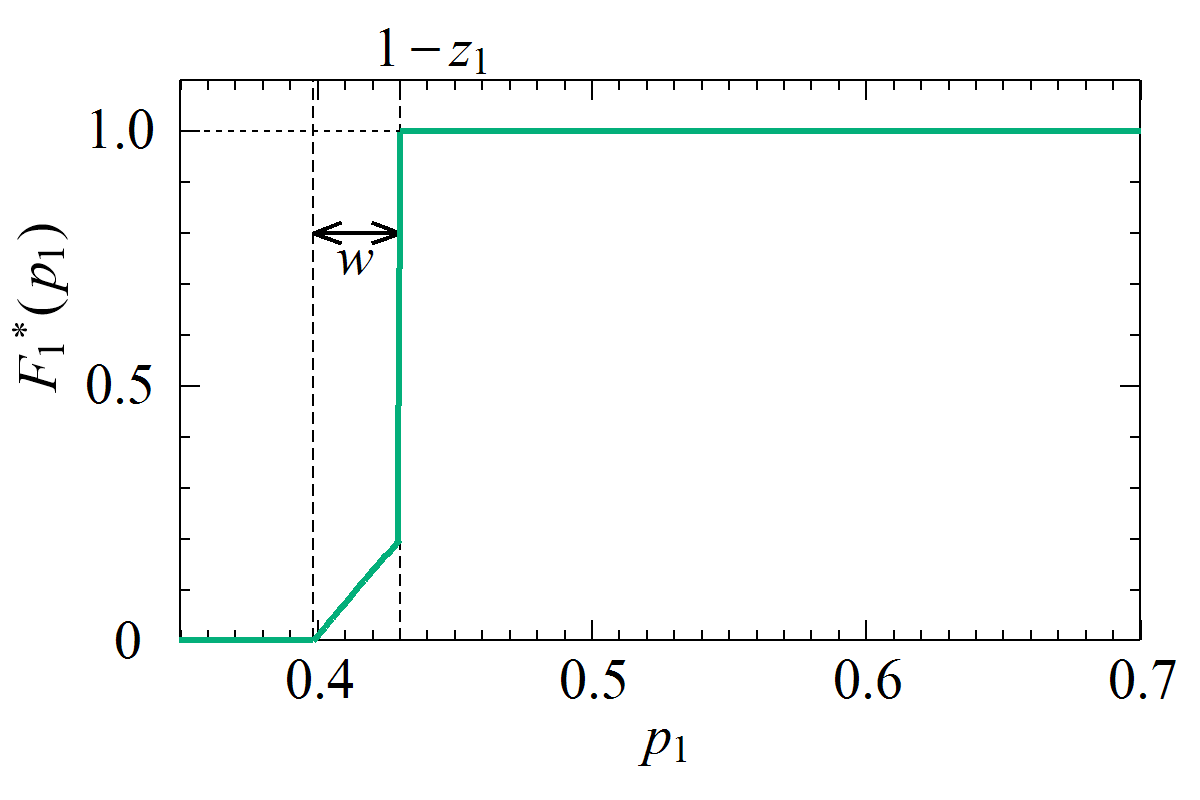}
\includegraphics[width=6cm]{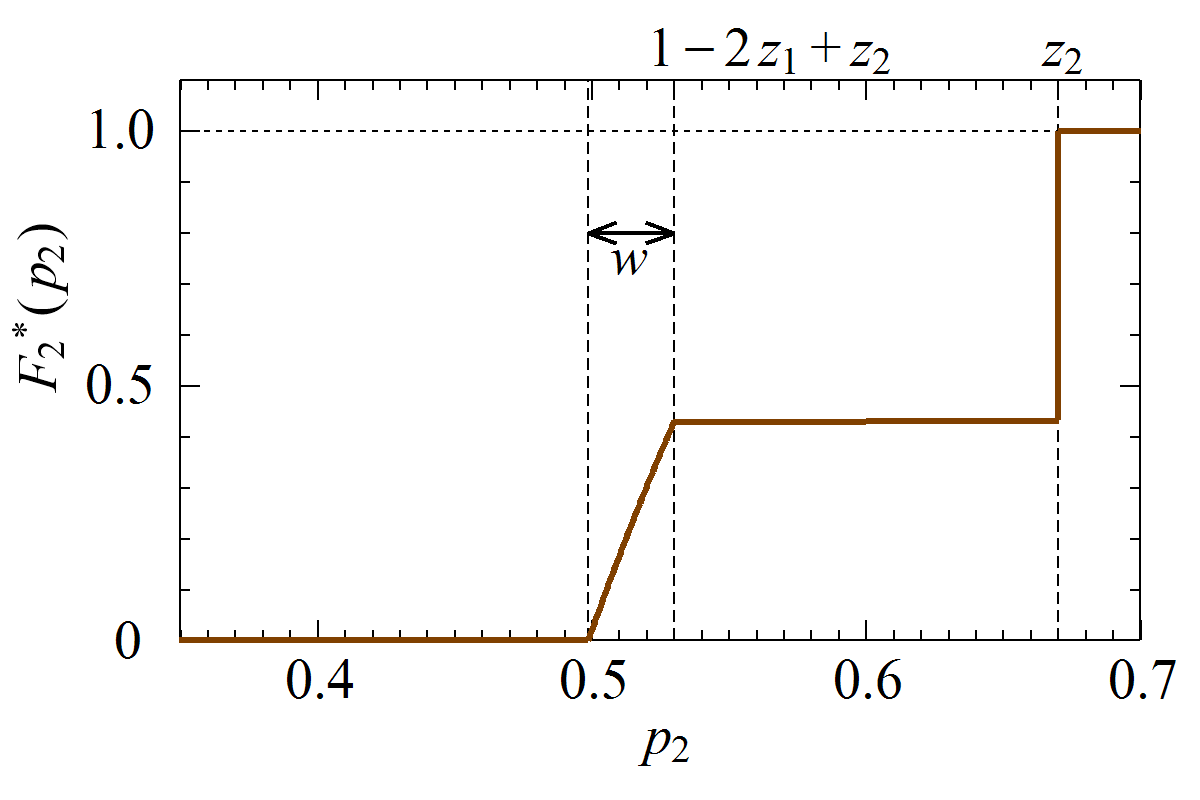}
\caption{Cumulative price distribution functions $F_1^\star$ (left) and $F_2^\star$ (right) for the mixed strategy equilibrium M3, drawn for $(z_1,z_2)=(0.57, 0.67)$.}
\label{fig:CDF_M3}
\end{figure}      

\section{Discussion}
In this section, we examine how the presence of captive buyers affects price competition in spatial competition. When Firm $1$ successfully takes all the informed consumers, Firm $2$ chooses to withdraw from the informed market and concentrate on maximizing profits from its captive buyers. Captive buyers play a significant role in this scenario. Furthermore, as will be discussed below, this behavior is common in both pure and mixed strategy equilibria. We characterized the equilibrium in which such behavior occurs.

First, we discuss pure strategy equilibria. A typical pure strategy equilibrium occurs when two firms are sufficiently distant from each other. Propositions~\ref{case1characterized1} and ~\ref{case1characterized2} are similar to the equilibrium identified by \citet{da} and \citet{op} in spatial competition without captive buyers. Proposition~\ref{case1characterized2} follows from the assumption that informed consumers have a limited reservation utility. Additionally, we have identified another pure strategy equilibrium. Propositions \ref{pure4-1} and \ref{pure4-2} demonstrate a pure strategy equilibrium when Firm $1$ is located further to the right of the center. In this case, a pure strategy equilibrium is obtained even if the firms are close to each other. Firm $2$ completely exits the informed market.

Second, we discuss mixed strategies. Mixed strategy equilibria emerge when $z_1$ is close to the center ($\because z_1+z_2 \ge 1$), where Firm $1$ has an advantage over Firm $2$ in price competition in the informed market. When one of the firms is near the center of the line segment (the median of the informed consumer), both firms are forced to adopt a mixed strategy. We call each interval with non-zero probability density an `island.' An island emerges when a firm faces different prices when targeting only its captive buyers versus when it gives weight to the chance to attract informed consumers. The atom of the equilibrium distribution $F_i$ appears at the maximum price that the firm can charge its captive buyers.

\begin{figure}[ht]
\centering
\includegraphics[width=6cm]{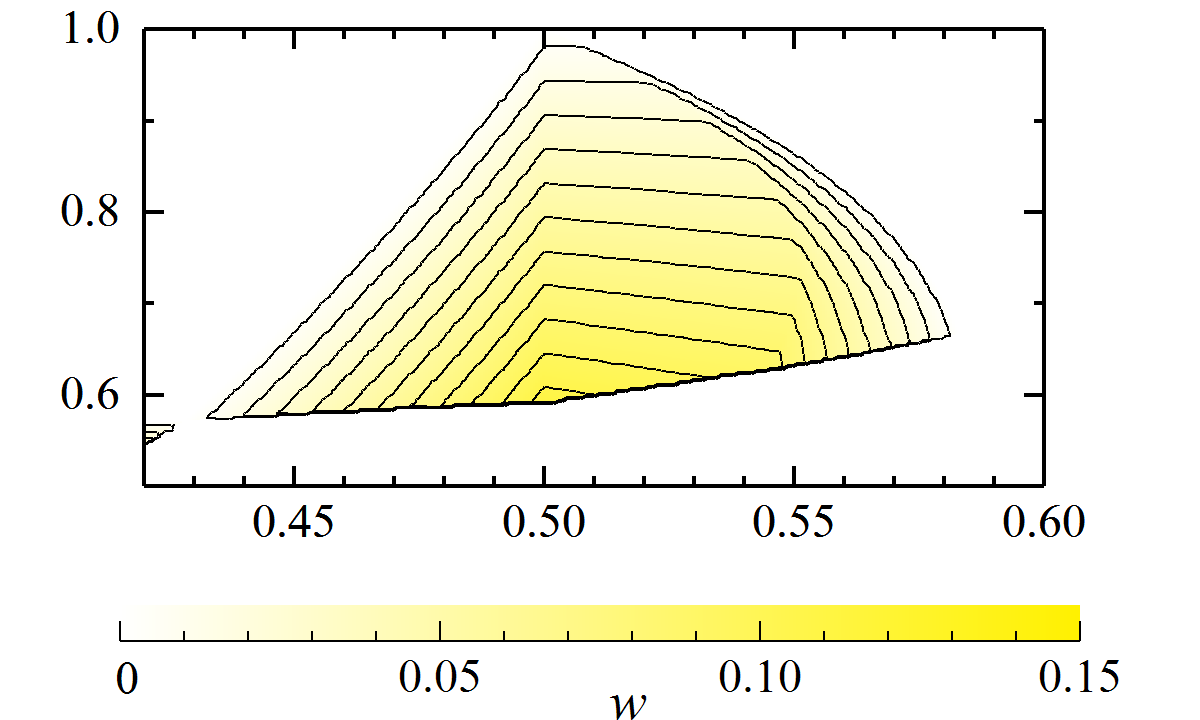}
\caption{The width $w$ of the equilibrium distributions is plotted as a function of $(z_1,z_2)$.
The contours are spaced by 0.01.}
\label{fig:w}
\end{figure}

We note the presence of a width, denoted as $w$, that is relevant to the price support in mixed strategy equilibria. $w$ is equal for both firms. The value of $w$ represents the margin that a firm has for price undercutting within the equilibrium. Figure~\ref{fig:w} plots the width $w$ of the equilibrium distributions as a function of $(z_1,z_2)$.

Third, we discuss the upper bound on equilibrium prices. The upper bound prices in this model, including the pure strategy equilibrium, are identical in all but one equilibrium. In our analysis, the upper bound of the interval typically represents the maximum price that both firms can extract from captive buyers. Firm $1$ is in a relatively advantageous position, and Firm $2$ is in a disadvantageous position. Firm $2$ usually sets its own reservation price for captive buyers, so the maximum price it can set is $p_2=z_2$. However, only mixed strategy equilibrium 2 is different. The price ceiling of Firm $2$ is at $1-2z_1+z_2$, which is lower than $z_2$ ($\because 1-2z_1<0$). In this case, the location of Firm $1$ affects the upper price limit of Firm $2$ because Firm $2$ benefits from lowering the price to obtain profits from the informed consumers rather than charging the maximum on its captive buyers. In other words, Firm $2$ takes the risk of lowering its price to attract informed consumers.
 
From the point of view of the form of the equilibrium pricing strategy and the region on the $z_1,z_2$ plane where it is realized, the property of the mixed strategy equilibria (M1, M2, and M3) is roughly similar to the equilibrium (T2) shown by the Osborne and Pitchik (OP) model.

Finally, we compare the profits of informed consumers and examine the differences between our model and the OP model. The key distinction lies in the presence of captive buyers at both ends of the line. In a typical spatial competition model, firms are located far from each other to avoid pressure in price competition. This implication is interpreted as maximizing product differentiation.

\begin{figure}[ht]
\centering
\includegraphics[width=6cm]{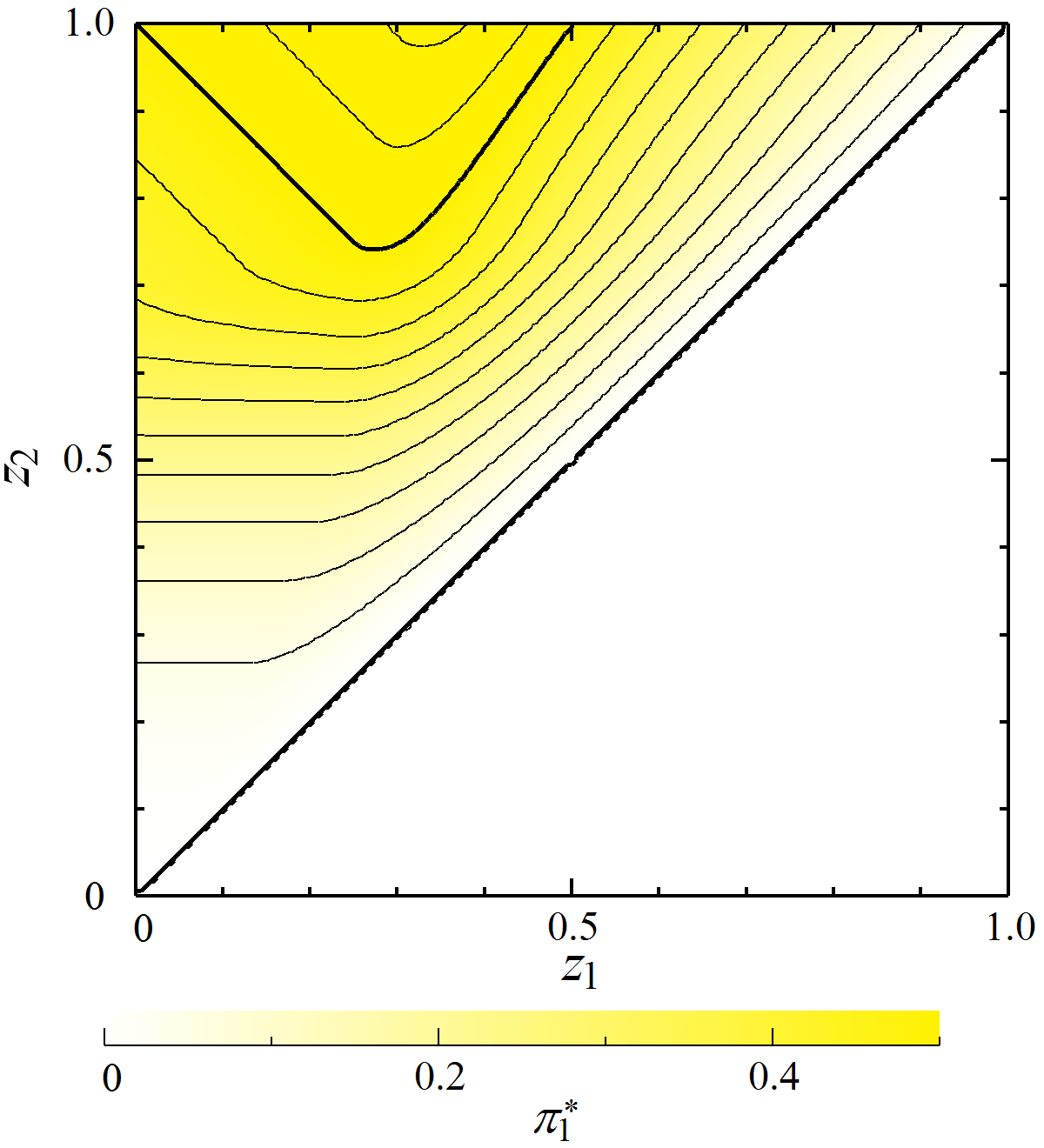}
\includegraphics[width=6cm]{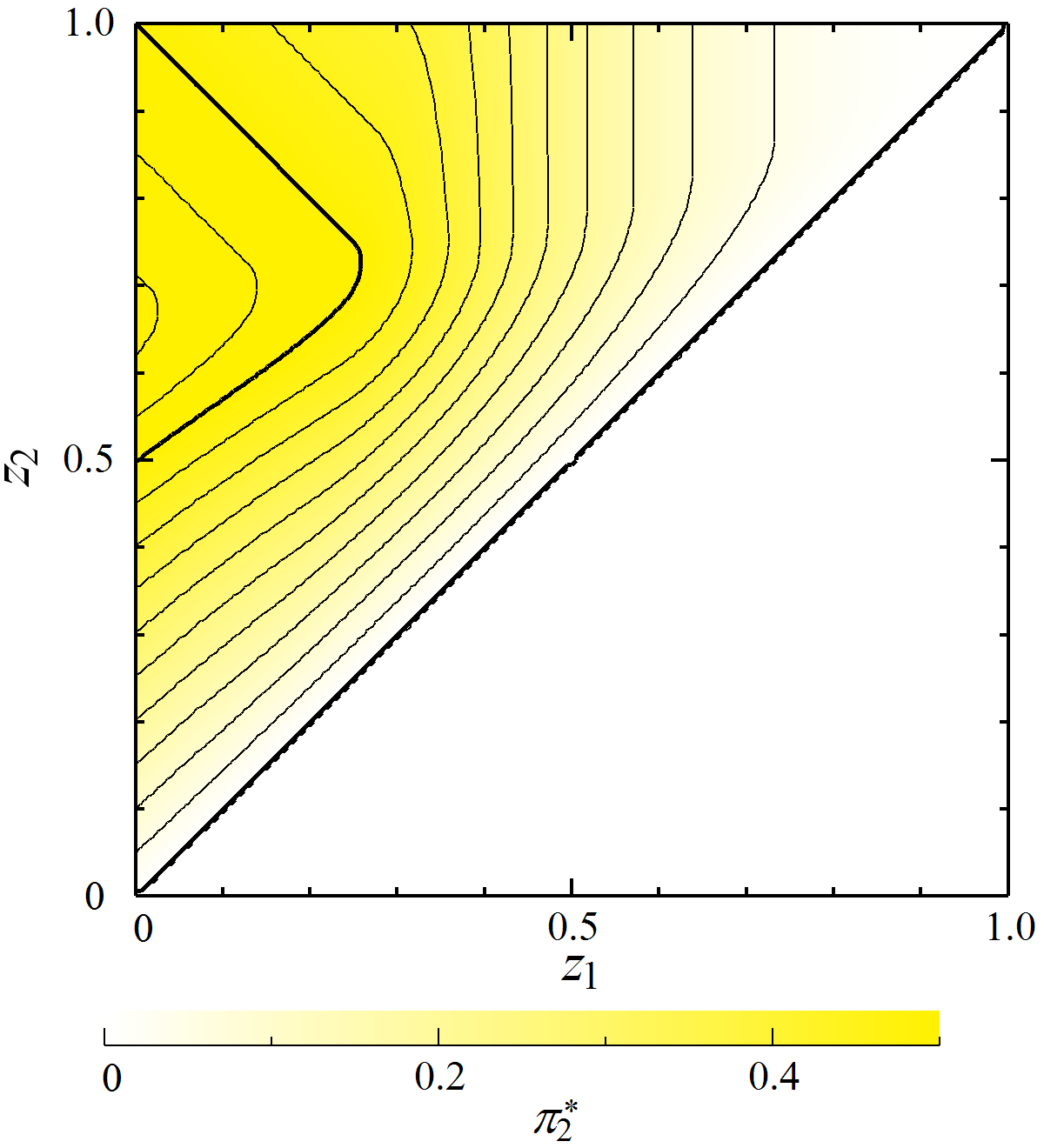}
\caption{Equilibrium profits $\pi_1^\star$ and $\pi_2^\star$ of the OP model are plotted as functions of $(z_1,z_2)$.
The contours are spaced by 0.1. The thick contours are drawn at $\pi_j^\star = 0.5$ and $1$.}
\label{fig:eqpiop}
\end{figure}

Figure~\ref{fig:eqpiop} reproduces the results of OP and shows the expected profits of each firm in equilibrium. The dark areas indicate location pairs with high profits. We can see that the upper left area is darker where the two firms are located far apart. Therefore, the result of OP also implies the maximum product differentiation. 

\begin{figure}[ht]
\centering
\includegraphics[width=6cm]{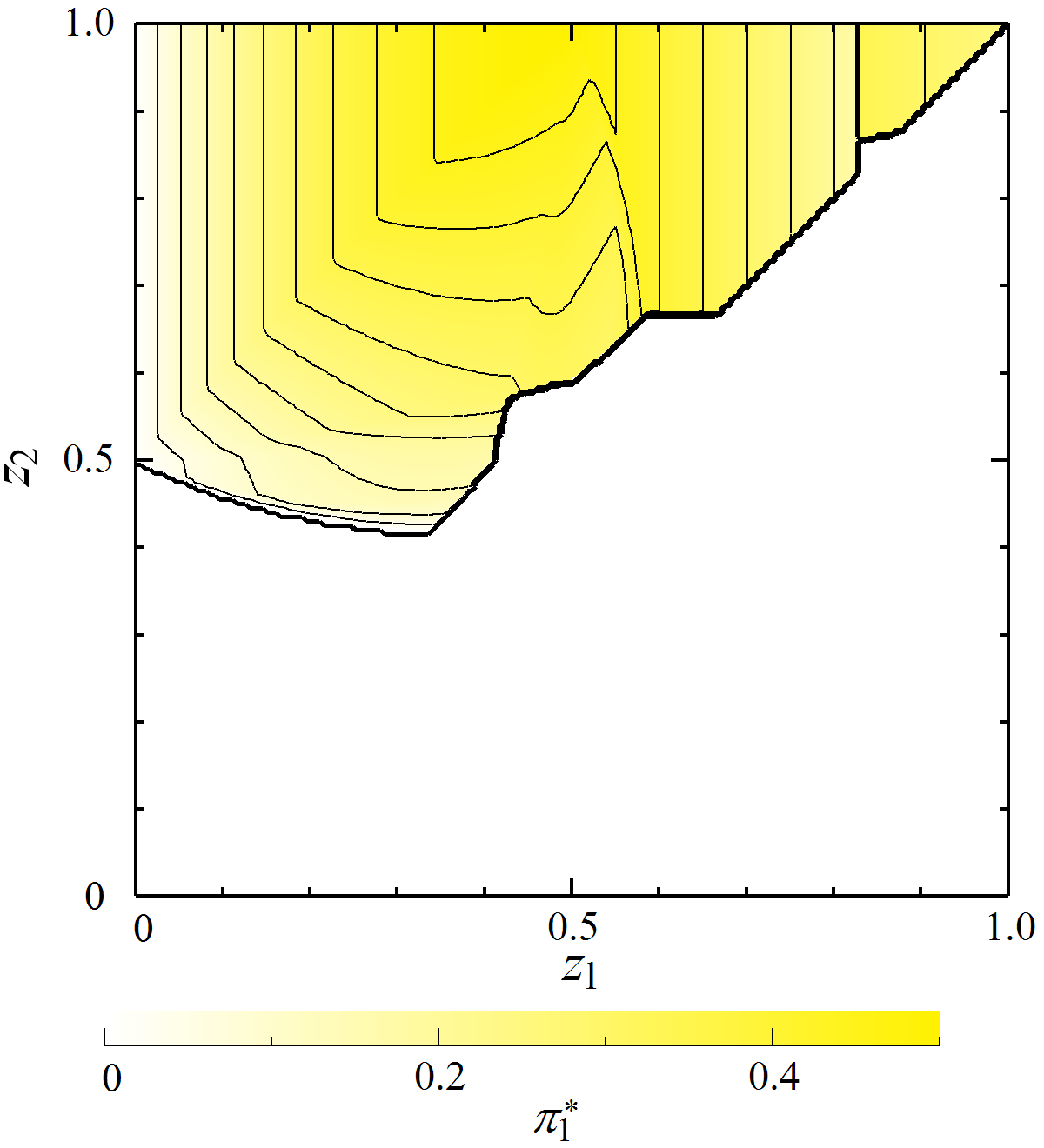}
\includegraphics[width=6cm]{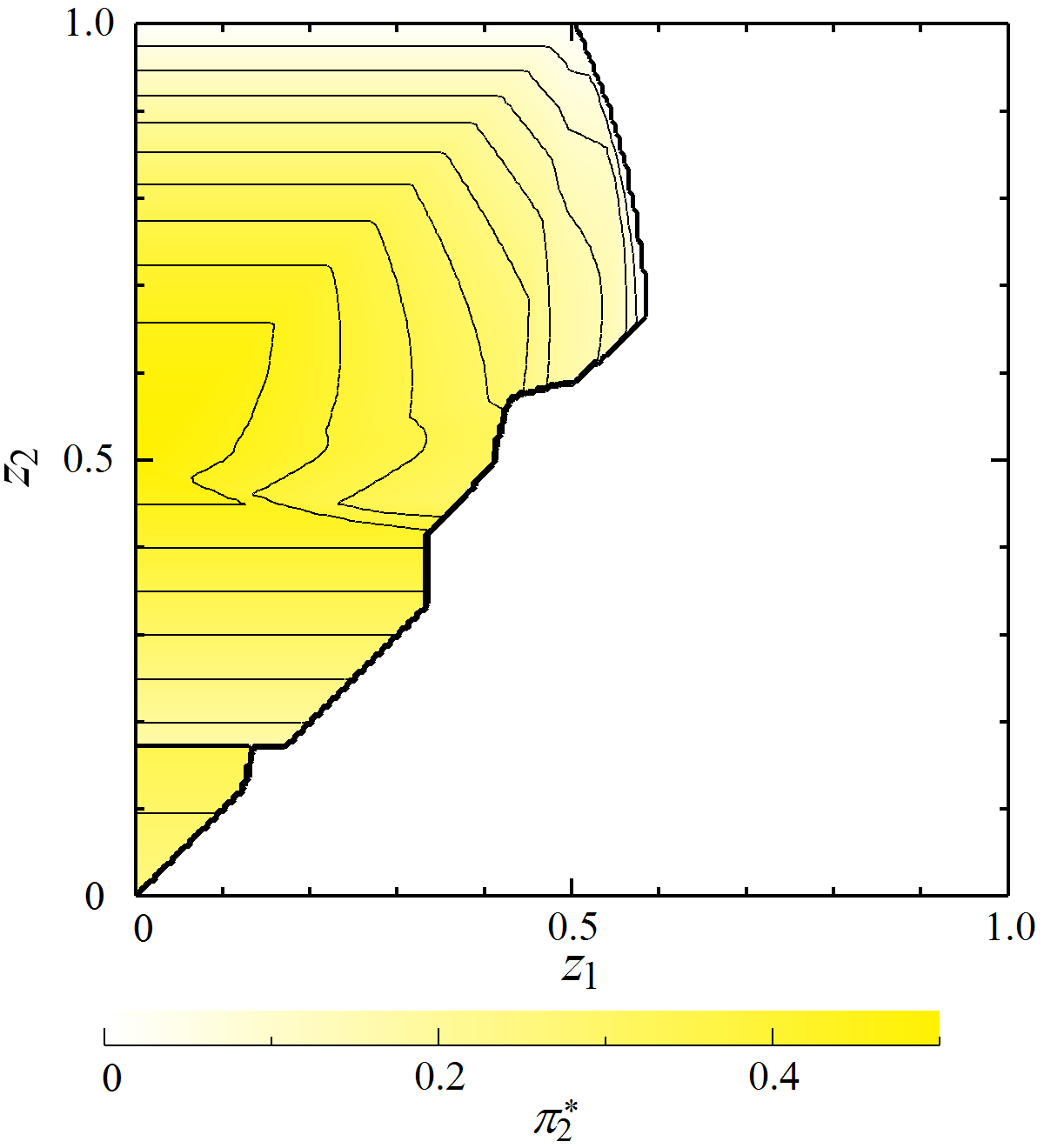}
\caption{Equilibrium profits $\pi_1^\star$ and $\pi_2^\star$ of our model are plotted as functions of $(z_1,z_2)$.
The contours are spaced by 0.1. The thick contours are drawn at $\pi_j^\star = 0.5$ and $1$.}
\label{fig:eqpikn}
\end{figure}

Figure~\ref{fig:eqpikn} illustrates the expected profits of our model. Firm $1$'s profit from informed consumers reaches its maximum near the center. If Firm $1$ is located further to the right of the center, Firm $2$ completely exits the informed market. Thus, Firm $2$'s profit is zero. The expected profit from the informed consumers is zero at the edge because if a firm is at the edge, it only focuses on the profit from its own captive buyer and does not expect profit from the informed consumer. 

\begin{figure}[ht]
\centering
\includegraphics[width=6cm]{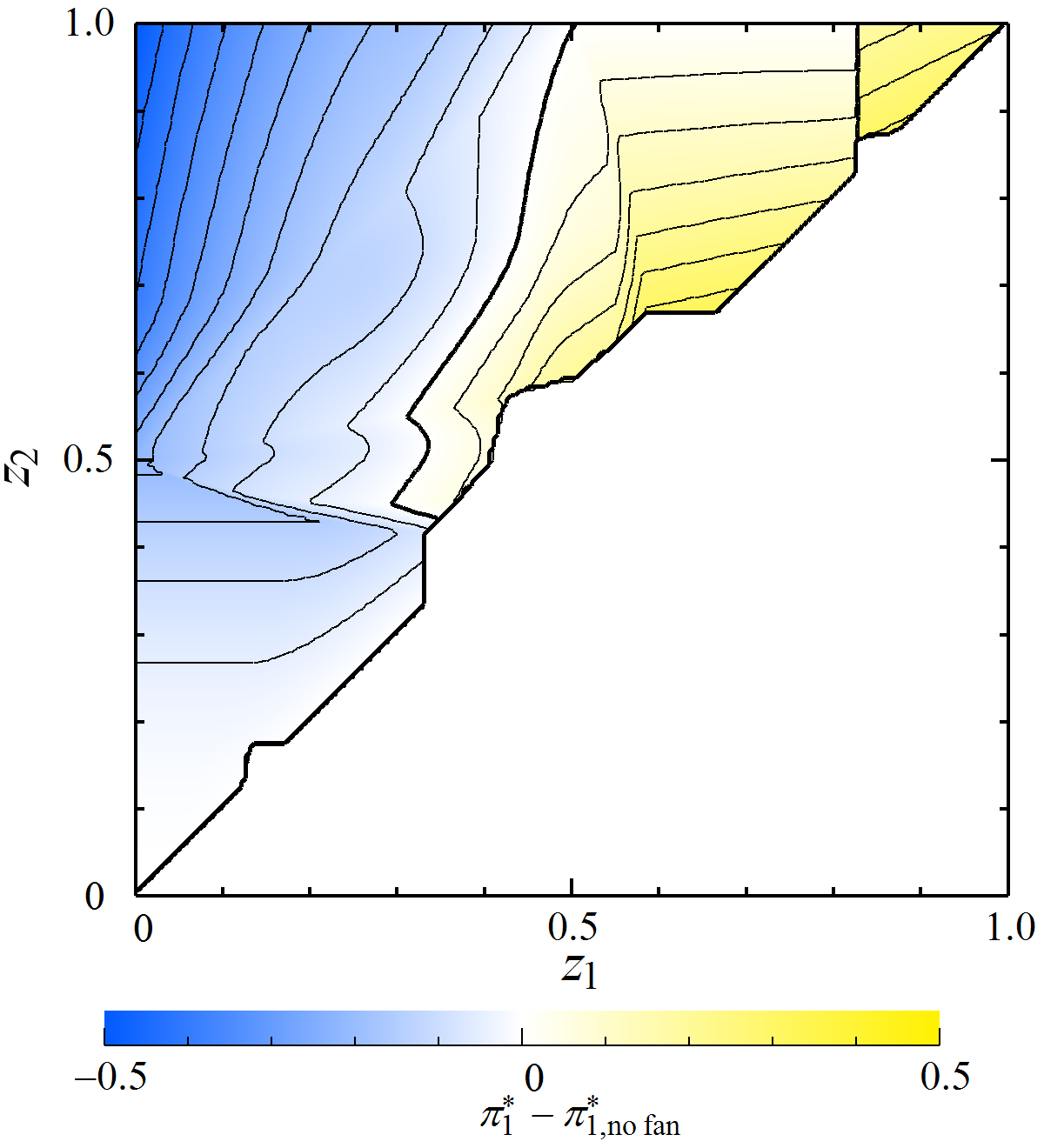}
\includegraphics[width=6cm]{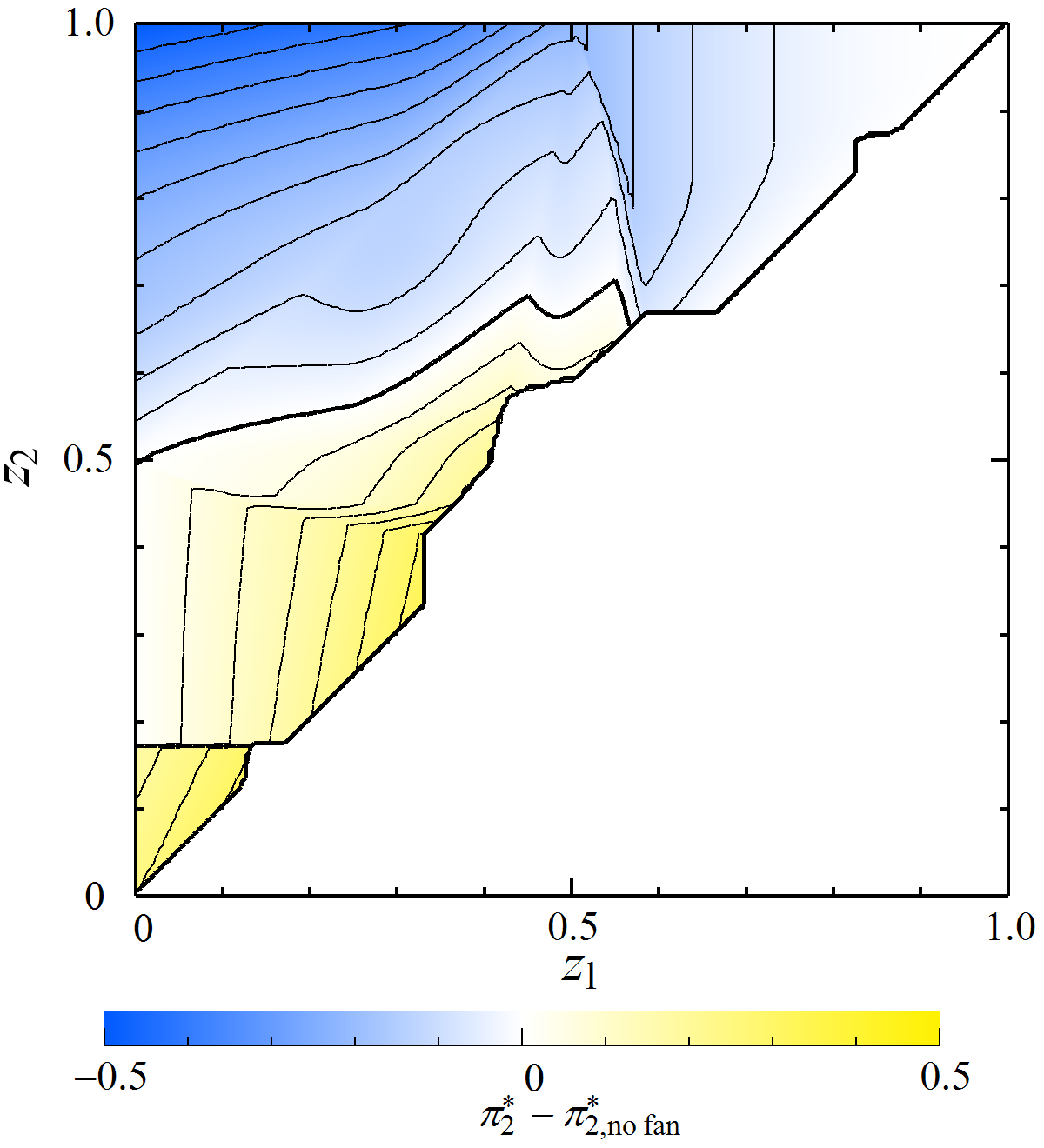}
\caption{The differences in equilibrium profits from informed consumers between the two models for each firm. In the yellow area, firms' equilibrium profits are higher in our model than in the OP model.
The contours are spaced by 0.1. The thick contours are drawn at $\pi_j^\star = 0.5$ and $1$.
}
\label{fig:bothpidiff}
\end{figure}

Figure~\ref{fig:bothpidiff} illustrates the difference in equilibrium profits between the OP model and our model. The above discussion can be seen more clearly in this figure. In the yellow zone, firms earn higher expected profits compared to the OP model. When both firms are beyond the center and relatively distant from their captive buyers, both firms' profits are higher than those of the OP model. In addition, Firm 1's yellow color becomes darker as it moves to the right because Firm 2 withdraws. In this case, Firm 1 is no longer in competition with Firm 2. The same thing can be observed in the lower left area for Firm 2, so this holds for both firms.

In all the mixed strategy equilibria we obtained, a mixed strategy equilibrium emerges as one firm moves closer to the center. The mixed strategy equilibrium, where both firms are near the center, is distinct from the equilibrium analyzed in this paper. When both firms are located close to one another, particularly near the center of the market, a firm that once lost informed consumers due to a competitor’s aggressive price cut can still attract them by lowering its prices again. Preliminary numerical calculations show intricate equilibrium, including an equilibrium where price supports split into multiple islands when firms are located closely at the center. 

The emergence of multiple islands implies that firms with captive buyers need more complex pricing strategies in their product differentiation competition when consumers perceive the product differentiation as relatively similar. \citeauthor{op} have reported a similar equilibrium (T1) in a spatial competition model without captive buyers. The T1 equilibrium reported by \citeauthor{op} aligns with our preliminary numerical calculations. However, the increased complexity of our model, due to the incorporation of captive buyers, leads to a region around the center that remains to be solved.

\section{Conclusion}
In this paper, we analyze the market with firms' product differentiation, taking into account the differences between "informed" and "uninformed" buyers of firms' prices and locations. We refer to the latter as "captive" buyers.

The existence of captive buyers leads to a situation where, in pure strategy equilibria, one firm captures informed consumers, while the other depends on captive buyers. Mixed strategy equilibria occur when one firm is closer to the center of the line segment. Typically, the upper-bound price in these equilibria reflects captive buyers’ reservation prices. However, in one mixed strategy equilibrium, Firm $2$ chooses to reduce its price to attract informed consumers rather than charging its captive buyers their maximum price. The equilibrium price support width, denoted as $w$, plays a critical role in price undercutting strategies.

Our model is based on \citeauthor{ho}'s model of spatial competition. Typically, consumers in \citeauthor{ho}'s model have information about prices and distances to all sellers, and they choose the seller with the lowest total price, including transportation costs. However, unlike this complete information setting, our model assumes that some consumers are uninformed about some sellers and buy from a particular seller. In this case, this seller has market power because it can secure profits from the "captive" buyers. Our model assumes that they exist at the ends of the line. This setting is the same as \citeauthor{nakagawa}'s model. 

Our study extends \citeauthor{nakagawa}’s model by uniformly distributing informed consumers and adopting linear transportation costs. This approach facilitates comparisons with standard spatial competition models, such as \citeauthor{op}. Comparing our model with \citeauthor{op}’s spatial competition model reveals distinctions stemming from captive buyers. The paper’s contributions include insights into pure and mixed strategy equilibria and the implications of captive buyers on competition. However, the unresolved issue remains the analysis of price competition in the market for slightly differentiated products. Furthermore, we cannot ensure that these equilibria are unique to each area, nor can we determine that the equilibrium payoffs are unique. Future research must address the challenge of fully understanding the complex interactions in the market for slightly differentiated products within spatial competition frameworks.

%%%%%%%%%%%%%%%%%%%%%%%%%%%%%%%%%%%%%%%%%%%%%%
%% Single Appendix:            %%
%%%%%%%%%%%%%%%%%%%%%%%%%%%%%%%%%%%%%%%%%%%%%%
%\begin{appendix}
%\section*{???} %% if no title is needed, leave empty \section*{}.
%\end{appendix}
%%%%%%%%%%%%%%%%%%%%%%%%%%%%%%%%%%%%%%%%%%%%%%
%% Multiple Appendixes:        %%
%%%%%%%%%%%%%%%%%%%%%%%%%%%%%%%%%%%%%%%%%%%%%%
\begin{appendix}

\section{Mathematical appendix for proofs}

\subsection{Proofs for Lemma~\ref{prelemmaofcondition3} and \ref{lemmaofcondition3}}\label{lemmaofcondition3appendix}
\begin{proof}[Proof for Lemma\ref{prelemmaofcondition3}]
Now, we check the condition when the informed consumer with their ideal point $t=z_1$ purchases $z_1$. From $p_1+z_1-z_1 < p_2 + \mid z_1-z_2\mid$ it follows that $p_1-p_2+z_1-z_2<0$. Similarly, when the informed consumer $t=z_2$ purchases $z_2$, we obtain $p_1-p_2+z_2-z_1>0$ by $p_1+\mid z_2-z_1 \mid< p_2 + z_2-z_2$. These two equations do not hold simultaneously if $z_1=z_2$. Thus $z_1 \ne z_2$.
\end{proof}

\begin{proof}[Proof for Lemma\ref{lemmaofcondition3}]
Suppose an informed consumer at $t \ne z_2$ is indifferent between $z_1$ and $z_2$. Then it follows that $p_1+|t-z_1| = p_2+|t-z_2|$ hold for $t \ne z_2$. Now, we verify this by considering the following three cases. 1, if $z_1 < z_2 \le t$ holds, we obtain that $p_1+|t-z_1|-p_2-|t-z_2|=p_1-p_2+t-z_1-t+z_2=p_1-p_2+z_2-z_1=0$. Thus $p_1+|t-z_1| = p_2+|t-z_2|$ hold in this case. In this case, our lemma holds. 2, if $t \le z_1 < z_2$ holds, we obtain that $p_1-p_2+z_1-z_2=0$. Thus, by $z_1 < z_2$, $p_1+|t-z_1| = p_2+|t-z_2|$ does not hold in this case. In this case, our lemma does not hold. 3, if $z_1 < t < z_2$, by $p_1+t-z_1=p_2+z_2-t$, we obtain that $p_1-p_2-z_1-z_2+2t=0$. Substituting $p_1-p_2=z_1-z_2$ into the \Eq{condition3}, we obtain that $z_1-z_2-z_1-z_2+2t=0$, therefore we have $t=z_2$. Thus, from cases 1 and 3, we obtain this lemma.
\end{proof}

\subsection{Proof for Proposition~\ref{allshopperz1}}\label{allshopperz1appendix}
\begin{proof}
First, we consider the condition that even the informed consumer with ideal point $t=z_2$ purchases the Firm $1$'s product with a characteristic $z_1$. This condition can be equivalently stated in the following equation as
\begin{equation}
p_1 + z_2 - z_1 < p_2 + z_2 - z_2,
\end{equation}
Here we obtain
\begin{equation}\label{condition1}
p_1 - p_2 + z_2 - z_1 < 0.
\end{equation}
Suppose for the sake of contradiction that an informed consumer with $t \ne z_2$ purchases the Firm $2$'s product with $z_2$ if \Eq{condition1} holds. Then, there exists $t$ such that
\begin{equation}\label{condition2}
p_1+|t-z_1| > p_2+|t-z_2|
\end{equation}
holds. This is a contradiction. Now consider the following three cases. 1, when $z_1 < z_2 \le t$, from \Eq{condition2} it follows that $p_1+|t-z_1|-p_2-|t-z_2|=p_1-p_2+t-z_1-t+z_2=p_1-p_2+z_2-z_1>0$. Thus, we have a contradiction. 2, when $t \le z_1 < z_2$, from \Eq{condition2} it follows that $p_1-p_2+z_1-z_2>0$. A contradiction. 3, when $z_1 < t < z_2$, since \Eq{condition1} holds, from \Eq{condition2} it follows that $p_1+|t-z_1|-p_2-|t-z_2|=p_1-p_2+t-z_1+z_2-t=p_1-p_2-z_1-z_2+2t \le p_1-p_2-z_1-z_2+2z_2=p_1-p_2-z_1+z_2<0$. A contradiction. And since we arrived at the contradiction, our original supposition that if \Eq{condition1} holds, there exists an informed consumer $t \ne z_2$ who purchases the Firm $2$'s product with $z_2$ can not be true. Therefore, if \Eq{condition1} holds, all informed consumers purchase the Firm $1$'s product.
\end{proof}

\subsection{Proof for Proposition~\ref{case1characterized1}}\label{case1characterized1appendix}
\begin{proof}
In this equilibrium, Firm $1$ obtains $\pi_1=(1-z_1)(\frac{1}{2}+z_1+z_2)$. Thus, we have the following equation \Eq{psubcond2-1} that Firm $1$ has no incentive to cut their price drastically to obtain all the informed consumers;
\begin{equation}
(1-z_1)(\frac{1}{2}+z_1+z_2) \ge 2z_1.
\end{equation}
Similarly, we have the following equation for Firm $2$;
\begin{equation}
z_2(\frac{5}{2}-z_1-z_2) \ge 2(1-z_2).
\end{equation}
Thus, we have $(p_1^{\star},p_2^{\star})=(1-z_1,z_2)$ if \Eq{psubcond2-1} and \Eq{psubcond2-2} hold. We also obtain that \Eq{psubcond2-1} and \Eq{psubcond2-2} hold if they charge a price pair $(p_1^{\star},p_2^{\star})=(1-z_1,z_2)$.
\end{proof}

\subsection{Proof for Proposition~\ref{case1characterized2}}\label{case1characterized2appendix}
\begin{proof}
From $z_2 \ge z_1 + \frac{1}{2}$, there is no $t$ satisfying $p_1+|t-z_1| = p_2+|t-z_2|$. Therefore, given $p_2=z_2$, it is sufficient to show that $\pi_1$ increases monotonically with respect to $p_1$ in the range satisfying $2+z_1-z_2 < p_1 <1-z_1$. We consider the condition for an informed consumer located to the right of Firm $1$ to purchase $z_1$: $p_1 \le 1-p_1$ ($\because$ the reservation value of consumers is $1$.). Informed consumers who purchase $z_1$ in the range are located in an interval of length $1-p_1$ rather than $t-z_1$. Therefore, the total number of informed consumers who buy $z_1$ is $z_1+1-p$, resulting in $\pi_1=p_1(1+z_1+1-p_1)$. The profit function is maximized at $1+{z_1}/2$. Thus, from $1-z_1 < 1+z_1/2$, $\pi_1$ is increasing monotonically with respect to $p_1$ in the range we are considering. In addition, in the range where $2+z_1-z_2 < p_1$, there exists $t$ satisfying $p_1+|t-z_1| = p_2+|t-z_2|$. Therefore, among the informed consumers located to the right of Firm $1$, the number of consumers who purchase $z_1$ is reduced below $1-p_1$ due to the competitive pressure from Firm $2$, which, together with the lower price itself, reduces its profit. 
\end{proof}

\subsection{Proof for Proposition~\ref{pure4-1}}\label{pure4-1appendix}
\begin{proof}
We show no unilateral deviation from $(p_1^{\star}, p_2^{\star})=(1-z_1, z_2)$. We will demonstrate this for the firm $2$ in the beginning. First, given $p_1^{\star}=1-z_1$, any deviation to $p_2 > p_2^{\star}$ results in zero profit. So, there is no deviation to $p_2 > p_2^\star$. Next, we check for deviations to $p_2 < p_2^{\star}$. Given $p_1^{\star}=1-z_1$ for Firm $1$, we must check for two conditions that Firm $2$ has no incentive to 1, win informed consumers on the interval $[z_2,1)$ by cutting its price to $1-z_1+(z_2-z_1)$ and; 2, win informed consumers on the interval $(0,z_2)$ by cutting its price to $1-z_1-(z_2-z_1)$. From the former condition, we get $\left(1-z_1+(z_2-z_1)\right)(1+(1-z_2))=(1-2z_1+z_2)(2-z_2) \le z_2$. From the latter condition we get $\left(1-z_1-(z_2-z_1)\right)(1+1)=(1-z_2)(2) \le z_2$. Since $2/3 \le z_2$, it holds. Here, with respect to $1-z_1-(z_2-z_1)\le p_2 \le 1-z_1+(z_2-z_1)$, $\pi_2=p_2(5-2z_1-z_z-p_2)/2$ to $\partial \pi_2/\partial p_2 >0$. Hence, Firm $2$ does not deviate.

Next, we consider a deviation by Firm $1$. Given $p_2^{\star}=z_2$, if it deviates to $p_1 < p_1^\star$, its profit will not increase because it has already won all of the informed consumers and its own captive buyers. Furthermore, $p_1^{\star}$ is the reservation price of the uninformed. Therefore, any deviation from $p_1 < p_1^{\star}$ will always result in a lower profit. Now consider the case where $p_1 > p_1^{\star}$. In this case, the available price for the deviation should be $p_1 < z_2-(z_2-z_1)=z_1$. Since $1/2 < z_1$ by $\left(1-z_1+(z_2-z_1)\right)(1+(1-z_2))=(1-2z_1+z_2)(2-z_2) < z_2$, we have $1-z_1 < z_1$. At the same time, we can also see that Firm $1$ only profits from the informed market due to this deviation. We calculate the Firm $1$'s product profit from this deviation. From $p_1+(z_1-t)\le p_1$, we see that all informed in the range satisfying $t \ge p_1-(1-z_1)$ are obtained by Firm $1$. Thus, we obtain $\pi_1^{\star} = p_1(1-t)=p_1(2-z_1-p_1)$. Therefore, the maximum profit of this deviation is given by $p_1=(2-z_1)/2$, that is, $\pi_1=(2-z_1)^2/4$. Conversely, the condition that this deviation does not occur is $2(1-z_1)\ge(2-z_1)^2/4$. Solving this, we get $-\frac{1}{4}(z_1+2)^2+2 \ge 0$. It follows that $-2(\sqrt{2}-1)\le z_1 \le 2(\sqrt{2}-1)$. Since $z_1 \le 2(\sqrt{2}-1)$, it does not deviate.

In equilibrium, Firm $1$ earns $\pi_1^{\star}=p_1^{\star}(1+1)=2(1-z_1)$ because it has won all its uninformed and informed consumers. Firm $2$ only gains from its own uninformed, so $\pi_2^{\star}=p_2^{\star}=z_2$.
\end{proof}

\subsection{Proof for Proposition~\ref{pure4-2}}\label{pure4-2appendix}
\begin{proof}
Given $p_2=z_2$, Firm $1$'s price to win all informed consumers in $[z_1, 1)$ on the right side of $z_1$ is $p_1 < z_1$ since $z_2-(z_2-z_1)=z_1$. Consider an equilibrium such that firm 1 gets all informed consumers in $[t,1)$. Then the profit of Firm $1$ can be evaluated as $\pi_1(p_1)=p_1(1-t)=p_1(1-(p_1-(1-z_1)))$, and the profit is maximized when $p_1= 1-z_1/2 < z_1$. Therefore, the equilibrium profit of Firm $1$ is $(1-z_1/2)^2$.

Suppose $(p_1^{\star}, p_2^{\star})=(1-z_1/2, z_2)$ to be an equilibrium. Given $p_2^{\star}=z_2$, consider any deviation such that $p_1 \le 1-z_1$. In this case, $\pi_1 = 2p_1 \le 2(1-z_1)$. Now that $z_1 > 2(\sqrt{2}-1)$ holds, then $2(1-z_1) \le (1-z_1/2)^2$. Consider any deviation such that $p_1 > z_1$. In this case, the profit is decreased because informed consumers that can be obtained on both sides of $z_1$ is greatly reduced. Therefore, it will not deviate.

Given $p_1^{\star}=1-z_1/2$, there occurs no deviation to $p_2 > z_2$ because the profit of firm $2$ is $0$. Since \Eq{pure2condi} is satisfied when \Eq{pure1condi} holds, there will not occur a deviation to $p_2 \le 1-z_1 + (z_2-z_1)$. Consider a deviation to a price such that $1-z_1 + (z_2-z_1) < p_2 < z_2$. Deviating to a price such that $1-z_1/2 + (z_2-z_1) < p_2 < z_2$ always reduces the profit of firm $2$, so no deviation occurs. Consider a deviation to a price such that $1-z_1/2 - (z_2-z_1) < p_2 \le 1-z_1/2 + (z_2-z_1)$. With a marginal consumer $t=(z_1+z_2+p_2-p_1)/2$ between $z_1$ and $z_2$, we get $\pi_2=p_2(1+(1-t))=p_2(5-3z_1/2-z_2-p_2)/2$. Since $\partial \pi_2/\partial p_2 = (5-3z_1/2-z_2-2p_2)/2 > (1/2+3(1-z_1)/2+(1-z_2)+2(1-p_2))/2 > 0$, $\pi_2$ is monotonically increasing. By substituting $p_2=1-z_1/2 + (z_2-z_1)$, we evaluate the maximum profit to be gained by this deviation. Since $z_1 > 2(\sqrt{2}-1) > 4/5, z_1 < z_2$, we get
\begin{align}
\pi_2=z_2-\frac{5}{2}\left(z_1-\frac{4}{5}\right)-(z_2-z_1) < z_2 = \pi_2^{\star}
\end{align}
So we obtain that firm $2$ does not deviate.

Finally, consider the deviation to a price where $1-z_1 + (z_2-z_1) < p_2 \le 1-z_1/2 - (z_2-z_1)$. Under this condition, firm $2$ is more attractive to informed consumers than Firm $1$, while it protects its own captive buyers. Therefore, since $p_2 + (z_2-t)\le 1$, informed consumers in the range $t \ge p_2 - (1-z_2)$ will purchase the product of firm $2$. So we get $\pi_2=p_2(3-z_2-p_2)$. We also get $\partial \pi_2/\partial p_2 = 3 - z_2 - 2p_2$. Since $p_2 < 1-z_1/2-(z_2-z_1) = 1-z_2/2-(z_2-z_1)/2 \le 1-z_2/2$, we have $2p_2 \le 2-z_2$, then $\partial \pi_2/\partial p_2 > 0$, and therefore $\pi_2$ is monotonically increasing. Substituting $p_2=1-z_1/2 - (z_2-z_1)$, we evaluate the maximum profit to be gained by this deviation. From \Eq{pure2condi}, $\pi_2 < z_2 = \pi_2^{\star}$, so the deviation will not occur.
\end{proof}

% \section{Proofs for Mixed Strategy Equilibria}

\subsection{Proof for Proposition~\ref{thm:sol1}}\label{Mix1proof}
The following lemmas give the proof of Proposition~\ref{thm:sol1}.
\begin{lemma}
For the probability distribution given by \Eq{eq:F2sol1} and when $z_1-w< p_1 < z_1$,
\begin{align}
\mathrm{E}[\pi_1](p_1) = \pi_1^\star.
\end{align}
\end{lemma}
\begin{proof}
By \Eq{eq:F2sol1}, $F_2(p_2)$ has no atoms in $z_1-w+(z_2-z_1)< p_2 < z_1+(z_2-z_1)$. Therefore, if $z_1-w< p_1 < z_1$, then
\begin{align*}
\mathrm{E}[\pi_1](p_1) ={} & p_1 \left( 1 
+ \int_{z_2-w}^{p_1+\de} \left( \frac{z_1+z_2+p_2-p_1}{2} \right) \diff F_2(p_2) + \left( 1 - F_2\left( p_1+\de \right) \right)
\right) \cr
={} & p_1 \Bigl( 2 - F_2( p_1+\de )
+ \left[\left( \frac{z_1+z_2+p_2-p_1}{2} \right) F_2(p_2) \right]_{z_2-w}^{p_1+\de} - \int_{z_2-w}^{p_1+\de} \frac{1}{2} F_2(p_2) \thinspace \diff p_2 \Bigr) \cr & \quad \text{($\because$ Integration by parts)} \cr
={} & p_1 \left( 2 - (1-z_2) F_2( p_1+\de )
 -  \int_{z_2-w}^{p_1+(z_2-z_1)} \frac{1}{2} h(p_2; z_2-w) \thinspace \diff p_2  
\right) \cr 
={} & p_1 \Biggl( 2 
- (1-z_2) h(p_1+\de; z_2-w) \cr
& \phantom{p_1 \Biggl( } {} - \int_{z_2-w}^{p_1+\de} 
\left(
- \frac{1}{2\lmb} \frac{\diff}{\diff p} h(p_2; z_2-w)  +  \frac{ 2(z_1-w) } { (p-\de)^2 }
\right) \diff p_2  
\Biggr) \  (\because\ \mbox{\Eq{eq:hdiff}}) \cr
={} & p_1 \left( 2 
- (1-z_2) h(p_1+\de; z_2-w) + (1-z_2) h(p_1+\de; z_2-w) + \frac{2(z_1-w)}{ p_1 } - 2
\right) \cr
={} & 2(z_1-w)
\end{align*}
\end{proof}

\begin{lemma}
For the probability distribution given by \Eq{eq:F2sol1} and when $p_1=1-z_1$,
\begin{align}
\mathrm{E}[\pi_1](p_1) = \pi_1^\star.
\end{align}
\end{lemma}
\begin{proof}
By \Eq{eq:F2sol1}, $F_2(p_2)$ has an atom at $p_2=z_2$. Therefore,
\begin{align}
\mathrm{P}[p_2=z_2] = F_2(z_2+0) - F_2(z_2-0) = 1 - h(z_2; z_2-w)
\end{align}
\begin{align}
\mathrm{E}[\pi_1](1-z_1) ={} & (1-z_1) \Biggl( 1 
+ \int_{z_2-w}^{z_2} \left( \frac{z_1+z_2+p_2-(1-z_1)}{2} \right) \diff F_2(p_2) \cr
& \phantom{(1-z_1) \Biggl(} {} + \left( \frac{z_1+z_2+z_2-(1-z_1)}{2} \right) \left( 1 - h(z_2; z_2-w) \right)
\Biggr) \cr
={} & (1-z_1) \Biggl( 1 
+ \left( z_1+z_2 - \frac{1}{2} \right) h(z_2; z_2-w)
- \int_{z_2-w}^{z_2} \frac{1}{2} F_2(p_2) \thinspace \diff p_2 \cr
&\phantom{(1-z_1) \Biggl(} {} + \left( z_1+z_2 - \frac{1}{2} \right) \left( 1 - h(z_2; z_2-w) \right)
\Bigr) \cr
={} & (1-z_1) \left( z_1+z_2 + \frac{1}{2}
- \int_{z_2-w}^{z_2} \left(
- \frac{1}{2\lmb} \frac{\diff}{\diff p} h(p_2; z_2-w)  +  \frac{ 2(z_1-w) } { (p-\de)^2 }
\right) \diff p_2  
\right) \cr
={} & (1-z_1) \left( z_1+z_2 + \frac{1}{2} + 
\frac{1}{2\lambda_2} h(z_2; z_2-w)  
+ \frac{2(z_1-w)}{ z_1 } - 2
\right) \cr
={} & (1-z_1) \left( z_1+z_2 + \frac{1}{2} - \frac{2w}{ z_1 } 
+ \frac{1}{2\lambda_2} h(z_2; z_2-w)  
\right) \label{eq:Epi1p1sh1}
\end{align}
Now, by virtue of \Eq{eq:defw1}, we obtain $\mathrm{E}[\pi_1](1-z_1) =2(z_1-w)$.
\end{proof}

\begin{lemma}
For the probability distribution given by \Eq{eq:F1sol1} and when $z_2-w\le p_2 \le z_2$,
\begin{align}
\mathrm{E}[\pi_2](p_2) = \pi_2^\star.
\end{align}
\end{lemma}
\begin{proof}
By \Eq{eq:F1sol1}, $F_1(p_1)$ has an atom at $p_1=1-z_1$. Therefore,
\begin{align}
\mathrm{P}[p_1=1-z_1] = F_1(1 - z_1 + 0) - F_1(1 - z_1-0) = 1 - g(z_1; z_1, \pi_2^\star)
\end{align}
\begin{align*}
\mathrm{E}[\pi_2](p_2) ={} & p_2 \Biggl(
1 + \int_{p_2-\de}^{z_1} \left(1-\frac{z_1+z_2+p_2-p_1}{2}\right) \diff F_1(p_1) \cr 
&\phantom{p_2 \Biggl(} {}+ \left(1-\frac{z_1+z_2+p_2-(1-z_1)}{2}\right)\left( 1 - g(z_1; z_1, \pi_2^\star) \right)\Biggr) \cr
={} & p_2 \Biggl(
1 + \left(\frac{2-z_2-p_2}{2}\right) F_1(z_1)
- (1-z_2) F_1(p_2-\de) 
- \int_{p_2-\de}^{z_1} \frac{1}{2} F_1(p_1) \thinspace \diff p_1 \cr
&\phantom{p_2 \Biggl(} {} \phantom{p_2\Biggl(} + \left(\frac{3-2z_1-z_2-p_2}{2}\right)
\left( 1 - g(z_1; z_1, \pi_2^\star) \right)
\Biggr) \cr
={} & p_2 \Biggl(
\left(\frac{5-2z_1-z_2-p_2}{2}\right)
+ \left(\frac{-1+2z_1}{2}\right)g(z_1; z_1, \pi_2^\star)
- (1-z_2) g(p_2-\de; z_1, \pi_2^\star) \cr
&\phantom{p_2 \Biggl(} {} - \int_{p_2-\de}^{z_1} \frac{1}{2} g(p_1; z_1, \pi_2^\star) \thinspace \diff p_1 
\Biggr) \cr
={} & p_2 \Biggl( 
\left( \frac{5-2z_1-z_2-p_2}{2}\right)
+ \left(\frac{-1+2z_1}{2}\right) g(z_1; z_1, \pi_2^\star)
- (1-z_2) g(p_2-\de; z_1, \pi_2^\star)  \cr
& \phantom{p_2\Biggl(}
- \int_{p_2-\de}^{z_1} \left( 
\frac{1}{2\lambda_2} \frac{\diff}{\diff p_1} g(p_1; z_1, \pi_2^\star)
+ \frac{1}{2} - \frac{\pi_2^\star}{ (p_1+\de)^2 }   \right) \thinspace \diff p_1 
\Biggr) \ (\because\ \mbox{\Eq{eq:gdiff}}) \cr
={} & p_2 \left( \frac{5}{2} - z_1 - z_2 
+ \left( z_1 + z_2 - \frac{3}{2} \right) g(z_1; z_1, \pi_2^\star) 
- \frac{\pi_2^\star}{z_2} + \frac{\pi_2^\star}{p_2} \right) \cr
={} & \pi_2^\star \ (\because\ \mbox{\Eq{eq:gbb}})
\end{align*}
\end{proof}

\begin{lemma}
For the probability distribution given by \Eq{eq:F1sol1} and when $p_2 = 1 - z_2$,
\begin{align}
\mathrm{E}[\pi_2](p_2-0) \leq \pi_2^\star. \label{eq:pi2bndM1}
\end{align}
\end{lemma}
\begin{proof}
\begin{align*}
\mathrm{E}[\pi_2](1-z_2-0) = & (1-z_2) 
 \left( 1 + \int_{z_1-w}^{z_1} \left(1-\frac{z_1+1-p_1}{2}\right) \diff F_1(p_1)  
 + \left( 1 - g(z_1;z_1,\pi_2^\star) \right)
 \right) \cr
 ={} &  (1-z_2) \left( 1 + \frac{1}{2} F_1(z_1) - \int_{z_1-w}^{z_1} \frac{1}{2} F_1(p_1) \thinspace \diff p_1 
 + \left( 1 - g(z_1;z_1,\pi_2^\star) \right)
 \right) \cr
 ={} &  (1-z_2) \left( 2 - \frac{1}{2} g(z_1;z_1,\pi_2^\star) - \int_{z_1-w}^{z_1} \frac{1}{2} g(p_1;z_1,\pi_2^\star) \thinspace \diff p_1 
 \right) \cr
 ={} &  (1-z_2) \Biggl( 2 - \frac{1}{2} g(z_1;z_1,\pi_2^\star) \cr
&\phantom{(1-z_2) \Biggl(} {} - \int_{z_1-w}^{z_1} \left( \frac{1}{2\lambda_2} \frac{\diff}{\diff p_1} g(p_1; z_1, \pi_2^\star) + \frac{1}{2} - \frac{\pi_2^\star}{ (p_1+\de)^2 }  \right) \thinspace \diff p_1  \Biggr) \cr
 ={} &  (1-z_2) \left( 2 - \left(\frac{3}{2}-z_2\right) g(z_1; z_1, \pi_2^\star) - \frac{w}{2} 
 - \frac{\pi_2^\star}{ z_2 } + \frac{\pi_2^\star}{ (z_2-w) }
 \right) \cr
 ={} &  (1-z_2) \Biggl( 2 - \frac{w}{2} - \left(\frac{3}{2}-z_2\right) g_0(z_1; z_1)  \cr
 & \phantom{(1-z_2) \Biggl(} {} +
 \left( \frac{1}{z_2-w} - \frac{1}{z_2} + \left(\frac{3}{2}-z_2\right) g_\pi(z_1; z_1) \right) \pi_2^\star
 \Biggr) \cr
 \leq {} & (1-z_2) \, \frac{\pi_2^\star}{1-z_2} \ (\because\ \mbox{\Eq{eq:condw1} and \Eq{eq:pi2e1}}) \cr
 = {} & \pi_2^\star
\end{align*}
\end{proof}

Strictly speaking, this is still not sufficient to prove that this price distribution is equilibrium. It is required to satisfy $\mathrm{E}[\pi_2](p_2)\leq \pi_2^\star$ for \emph{all} $p_2$, not only at $p_2=1-z_2$, so that Firm 2 has no incentive to deviate from this strategy. However, numerical verification showed that once \Eq{eq:pi2bndM1} holds, we have $\mathrm{E}[\pi_2](p_2)\leq \pi_2^\star$ for all the other $p_2$.

\subsection{Proof for Proposition~\ref{thm:sol2}}\label{Mix2proof}
The following lemmas give the proof of Proposition~\ref{thm:sol2}.
\begin{lemma}
For the probability distribution given by \Eq{eq:F2sol2} and when $1-z_1-w< p_1 < 1-z_1$,
\begin{align}
\mathrm{E}[\pi_1](p_1) = \pi_1^\star.
\end{align}
\end{lemma}
\begin{proof}
By \Eq{eq:F2sol2}, $F_2^\star(p_2)$ has no atoms in $1-z_1-w+(z_2-z_1)< p_2 < 1-z_1+(z_2-z_1)$. Therefore, when $1-z_1-w< p_1 < 1-z_1$,
\begin{align*}
\mathrm{E}[\pi_1](p_1) ={} & p_1 \left( 1 
+ \int_{1-2z_1+z_2-w}^{p_1+\de} \left( \frac{z_1+z_2+p_2-p_1}{2} \right) \thinspace \diff F_2^\star(p_2) + \left( 1 - F_2\left( p_1+\de \right) \right)
\right) \cr
={} & p_1 \biggl( 1 
+ \left[ \left( \frac{z_1+z_2+p_2-p_1}{2} \right) F_2^\star(p_2) \right]_{1-2z_1+z_2-w}^{p_1+\de} -  \int_{1-2z_1+z_2-w}^{p_1+\de} \frac{1}{2} F_2^\star(p_2) \thinspace \diff p_2 \cr
&\phantom{p_1 \bigg(} {} + 1 - F_2( p_1+\de ) 
\biggr)  \cr &  \text{($\because$ Integration by parts)} \cr
={} & p_1 \left( 2 
- (1-z_2) F_2(p_1+\de) - \int_{1-2z_1+z_2-w}^{p_1+\de} \frac{1}{2} h(p_2; 1-2z_1+z_2-w) \thinspace \diff p_2  ) 
\right) \cr
={} & p_1 \biggl( 2 
- (1-z_2) h(p_1+\de; 1-2z_1+z_2-w) \cr 
& \phantom{p_1 \biggl(} {}- \int_{1-2z_1+z_2-w}^{p_1+\de} 
\left( - \frac{1}{2\lambda_2} \frac{\diff}{\diff p_2} h(p_2; 1-2z_1+z_2-w)
+ \frac{2(1-z_1-w)}{ (p_2-\de)^2 }  \right) \thinspace \diff p_2  
\biggr) \cr & (\because\ \text{\Eq{eq:hdiff}}) \cr
={} & p_1 \biggl( 2 
- (1-z_2) h(p_1+\de; 1-2z_1+z_2-w) \cr 
& \phantom{p_1 \biggl(} {} +
\frac{1}{2\lambda_2} h(p_1+\de; 1-2z_1+z_2-w)
+ \frac{2(1-z_1-w)}{ p_1 } - 2 \biggr) \cr
={} & 2(1-z_1-w) \quad (\because \text{\Eq{eq:lmbdef}})
\end{align*}
\end{proof}

\begin{lemma}
For the probability distribution given by \Eq{eq:F1sol2} and when $1-2z_1+z_2-w < p_2 < 1-2z_1+z_2$,
\begin{align}
\mathrm{E}[\pi_2](p_2) = \pi_2^\star.
\label{eq:pi2const2}
\end{align}
\end{lemma}
\begin{proof}
By \Eq{eq:F1sol2}, $F_1^\star(p_1)$ has an atom at $p_1=1-z_1$. Therefore,
\begin{align}
\mathrm{P}[p_1=1-z_1] = 1 - F_1^\star(1-z_1-0) 
= 1 - g(1-z_1; 1-z_1, \pi_2^\star) 
\label{eq:F1b2}
\end{align}
\begin{align*}
\mathrm{E}[\pi_2](p_2) ={} & p_2 \Biggl(
1 + \int_{p_2-\de}^{1-z_1} \left(1-\frac{z_1+z_2+p_2-p_1}{2}\right) \thinspace \diff F_1^\star(p_1) \cr 
& \phantom{p_2\Biggl(} 
{} + \left(1-\frac{z_1+z_2+p_2-(1-z_1)}{2}\right)
\mathrm{P}[p_1=1-z_1]
\Biggr) \cr
={} & p_2 \Biggl(
1 + \left(\frac{3-2z_1-z_2-p_2}{2}\right) F_1^\star(1-z_1-0)
- (1-z_2) F_1^\star(p_2-\de) \cr 
& \phantom{p_2\Biggl(} 
{} 
- \int_{p_2-\de}^{1-z_1} \frac{1}{2} F_1^\star(p_1) \thinspace \diff p_1 
+ \left(\frac{3-2z_1-z_2-p_2}{2}\right)
\left( 1 - F_1^\star(1-z_1-0) \right)
\Biggr) \cr
={} & p_2 \Biggl(
\left(\frac{5-2z_1-z_2-p_2}{2}\right)
- (1-z_2) F_1^\star(p_2-\de) 
- \int_{p_2-\de}^{1-z_1} \frac{1}{2} F_1^\star(p_1) \thinspace \diff p_1 
\Biggr) \cr
={} & p_2 \Biggl(
\left(\frac{5-2z_1-z_2-p_2}{2}\right)
- (1-z_2) g(p_2-\de; 1-z_1, \pi_2^\star) \cr &
\phantom{p_2 \Biggl(} {} - \int_{p_2-\de}^{1-z_1} \frac{1}{2} g(p_1; 1-z_1, \pi_2^\star) \thinspace \diff p_1 
\Biggr) \cr
={} & p_2 \Biggl(
\left(\frac{5-2z_1-z_2-p_2}{2}\right)
- (1-z_2) g(p_2-\de; 1-z_1, \pi_2^\star) \cr
& \phantom{p_2\Biggl(}
- \int_{p_2-\de}^{1-z_1} \left( \frac{1}{2\lambda_2} \frac{\diff}{\diff p_1} g(p_1; 1-z_1, \pi_2^\star) + \frac{1}{2} - \frac{\pi_2^\star}{ (p_1+\de)^2 }  \right) \thinspace \diff p_1 
\Biggr) \quad (\because \text{\Eq{eq:gdiff}}) \cr
={} & p_2 \Biggl(
\left(\frac{5-2z_1-z_2-p_2}{2}\right)
- (1-z_2) g(p_2-\de; 1-z_1, \pi_2^\star) 
\cr
& \phantom{p_2\Biggl(} {}
- \frac{1}{2\lambda_2} g(1-z_1; 1-z_1, \pi_2^\star)
+ \frac{1}{2\lambda_2} g(p_2-\de; 1-z_1, \pi_2^\star)\cr
& \phantom{p_2\Biggl(} {}
- \frac{(1-z_1-p_2+\de)}{2} - \frac{\pi_2^\star}{ (1-z_1+\de)} +  \frac{\pi_2^\star}{ p_2 }   
\Biggr) \cr
={} & p_2 \left( 2 - z_2 
- \frac{\pi_2^\star}{1-2z_1+z_2} + \frac{\pi_2^\star}{p_2} 
-  \frac{1}{2\lambda_2} g(1-z_1; 1-z_1, \pi_2^\star)
\right) \cr
={} & p_2 \left( 
2 - z_2 
- \frac{\pi_2^\star}{1-2z_1+z_2} + \frac{\pi_2^\star}{p_2} 
- \frac{1}{2} \left( (4-2z_2) - \frac{2\pi_2^\star}{1-z_1+\de}  \right) \right) \quad(\because \text{\Eq{eq:gbb}})\cr
={} & \pi_2^\star
\end{align*}
\end{proof}

\begin{lemma}
For the probability distribution given by \Eq{eq:F1sol2} and when $p_2 = z_2-0$,
\begin{align}
\mathrm{E}[\pi_2](p_2) \leq \pi_2^\star. \label{eq:pi2bndM2a}
\end{align}
\end{lemma}
\begin{proof}
Due to \Eq{eq:F1sol2}, the price $p_1$ of Firm 1 is lower than $1-z_1$ with probability 1. The latter is lower than $z_2-\de$ because of \Eq{eq:dedef} and \Eq{eq:rgn2a}. Therefore, when $p_2=z_2$, Firm 2 gains only from its fan. Then,
\begin{align*}
\mathrm{E}[\pi_2](z_2-0) = z_2 \leq \pi_2^\star,
\end{align*}
owing to \Eq{eq:condw2b} and \Eq{eq:pi2e2}.
\end{proof}

\begin{lemma}
For the probability distribution given by \Eq{eq:F1sol2} and when $p_2 = 1 - z_2 -0$,
\begin{align}
\mathrm{E}[\pi_2](p_2) \leq \pi_2^\star. \label{eq:pi2bndM2b}
\end{align}
\end{lemma}
\begin{proof}
In this case, we have $p_2+\de = 1 - z_1 -0$. Therefore, when Firm 1 chooses $p_1=1-z_1$, Firm 2 gains all of the shoppers. 
\begin{align*}
\mathrm{E}[\pi_2](1-z_2-0) = {} & {} (1-z_2) 
\left( 1 + \int_{1-z_1-w}^{1-z_1} \left(1-\frac{z_1+1-p_1}{2}\right) \diff F_1^\star(p_1) 
+ \left( 1 - F_1^\star(1-z_1-0) \right)
\right) \cr
={} &  (1-z_2) \left( 2 - z_1 F_1^\star(1-z_1-0) - \int_{1-z_1-w}^{1-z_1} \frac{1}{2} F_1^\star(p_1) \thinspace \diff p_1 
\right) \cr
={} &  (1-z_2) \left( 2 - z_1 g(1-z_1; 1-z_1, \pi_2^\star) - \int_{1-z_1-w}^{1-z_1} \frac{1}{2} g(p_1; 1-z_1, \pi_2^\star) \thinspace \diff p_1 
\right) \cr
={} &  (1-z_2) \Biggl( 2 - z_1 g(1-z_1; 1-z_1, \pi_2^\star) 
\cr & \phantom{(1-z_2)\Biggl(} {} 
- \int_{1-z_1-w}^{1-z_1} \left( \frac{1}{2\lambda_2} \frac{\diff}{\diff p_1} g(p_1; 1-z_1, \pi_2^\star) +\frac{1}{2} - \frac{\pi_2^\star}{(p_1+\de)^2} \right) \thinspace \diff p_1 
\Biggr) \cr
={} &  (1-z_2) \Biggl( 2 - \left(1+z_1-z_2\right) g(1-z_1; 1-z_1, \pi_2^\star) - \frac{w}{2}
\cr & \phantom{(1-z_2)\Biggl(} {} 
- \frac{\pi_2^\star}{ (1-2z_1+z_2) } + \frac{\pi_2^\star}{  (1-2z_1+z_2-w) }
\Biggr) \cr
={} &  (1-z_2) \Biggl( 2 - \left(\frac{1+z_1-z_2}{2-2z_2}\right) \left( (4-2z_2) - \frac{2\pi_2^\star}{(1-2z_1+z_2)} \right) - \frac{w}{2}
\cr & \phantom{(1-z_2)\Biggl(} {} 
- \frac{\pi_2^\star}{ (1-2z_1+z_2) } + \frac{\pi_2^\star}{  (1-2z_1+z_2-w) }
\Biggr) \cr
={} &  \left( \frac{z_1}{(1-2z_1+z_2)} + \frac{(1-z_2)}{(1-2z_1+z_2-w)}\right) \pi_2^\star \cr
& - \left(z_1 - (1-z_2)(z_2-z_1)  + \frac{(1-z_2)w}{2} \right) \cr
\leq {} & \pi_2^\star,
\end{align*}
owing to \Eq{eq:condw2a} and \Eq{eq:pi2e2}.
\end{proof}

As was the case for the proof of Proposition~\ref{thm:sol1}, this is strictly not sufficient to prove that this price distribution is equilibrium. It is required to satisfy $\mathrm{E}[\pi_2](p_2)\leq \pi_2^\star$ for \emph{all} $p_2$, not only at $p_2=1-z_2$, so that Firm 2 has no incentive to deviate from this strategy. However, numerical verification showed that once \Eq{eq:pi2bndM2a} and \Eq{eq:pi2bndM2b} hold, we have $\mathrm{E}[\pi_2](p_2)\leq \pi_2^\star$ for all the other $p_2$.

\subsection{Proof for Proposition~\ref{thm:sol3}}\label{Mix3proof}
The following lemmas give the proof of Proposition~\ref{thm:sol3}.
\begin{lemma}
For the probability distribution given by \Eq{eq:F2sol3} and when $1-z_1-w< p_1 < 1-z_1$,
\begin{align}
\mathrm{E}[\pi_1](p_1) = \pi_1^\star
\end{align}
\end{lemma}
\begin{proof}
By \Eq{eq:F2sol3}, $F_2(p_2)$ has no atoms in $1-z_1-w+(z_2-z_1)< p_2 < 1-z_1+(z_2-z_1)$. Therefore, if $1-z_1-w< p_1 < 1-z_1$, then
\begin{align}
\mathrm{E}[\pi_1](p_1) ={} & p_1 \left( 1 
+ \int_{1-2z_1+z_2-w}^{p_1+\de} \left( \frac{z_1+z_2+p_2-p_1}{2} \right) \diff F_2^\star(p_2) + \left( 1 - F_2^\star \left( p_1+\de \right) \right)
\right) \cr
={} & p_1 \Biggl( 1 
+ \left[ \left( \frac{z_1+z_2+p_2-p_1}{2} \right) F_2^\star(p_2) \right]_{1-2z_1+z_2-w}^{p_1+\de} 
\cr & \phantom{p_1 \Biggl(} {}
-  \int_{1-2z_1+z_2-w}^{p_1+\de} \frac{1}{2} F_2^\star(p_2) \thinspace \diff p_2 + 1 - F_2^\star( p_1+\de ) 
\Biggr)  \cr & \text{($\because$ Integration by parts)} \cr
={} & p_1 \left( 1 
+ z_2 F_2^\star(p_1+\de) -  \int_{1-2z_1+z_2-w}^{p_1+\de} \frac{1}{2} F_2^\star(p_2) \thinspace \diff p_2 + 1 - F_2^\star( p_1+\de ) 
\right)\cr
={} & p_1 \biggl( 2 
- (1-z_2) h(p_1+\de; 1-2z_1+z_2-w) 
\cr & \phantom{p_1 \biggl(} {}
- \int_{1-2z_1+z_2-w}^{p_1+\de} \frac{1}{2} h(p_2; 1-2z_1+z_2-w) \thinspace \diff p_2  
\biggr) \cr
={} & p_1 \Biggl( 2 
- (1-z_2) h(p_1+\de; 1-2z_1+z_2-w) 
\cr & \phantom{p_1\Biggl(} {}- \int_{1-2z_1+z_2-w}^{p_1+\de} 
\left( - \frac{1}{2\lambda_2} \frac{\diff}{\diff p_2} h(p_2; 1-2z_1+z_2-w) + \frac{2(1-z_1-w)}{ (p_2-\de)^2 } \right) \thinspace \diff p_2  
\Biggr) 
\cr & \quad (\because \text{\Eq{eq:hdiff}}) \cr
={} & p_1 \Biggl( 2 
- (1-z_2) h(p_1+\de; 1-2z_1+z_2-w) 
\cr & \phantom{p_1 \Biggl(} {}
+ (1-z_2) h(p_1+\de; 1-2z_1+z_2-w)
+ \frac{2(1-z_1-w)}{ p_1 } - 2
\Biggr) \cr
={} & 2(1-z_1-w)
\end{align}
\end{proof}

\begin{lemma}
For the probability distribution given by \Eq{eq:F1sol3} and when $1-2z_1+z_2-w < p_2 < 1-2z_1+z_2$,
\begin{align}
\mathrm{E}[\pi_2](p_2) = \pi_2^\star
\label{eq:pi2const3}
\end{align}
\end{lemma}
\begin{proof}
By \Eq{eq:F1sol3}, $F_1^\star(p_1)$ has an atom at $p_1=1-z_1$. Therefore, by \Eq{eq:gbb} and \Eq{eq:pi2e3}
\begin{align}
\mathrm{P}[p_1=1-z_1] = 1 - F_1^\star(1-z_1-0) 
= 1 - g(1-z_1; 1-z_1, \pi_2^\star) = -2\lambda_2 + \frac{2\lambda_2 z_2 }{1-2z_1+z_2}. 
\label{eq:F1b3}
\end{align}
\begin{align*}
\mathrm{E}[\pi_2](p_2) ={} & p_2 \Biggl( 1 + \int_{p_2-\de}^{1-z_1} \left(1-\frac{z_1+z_2+p_2-p_1}{2}\right) \diff F_1^\star(p_1) 
\cr & \phantom{p_2\Biggl(} {}
+ \left(1-\frac{z_1+z_2+p_2-(1-z_1)}{2}\right)
\mathrm{P}[p_1=1-z_1]
\Biggr) \cr
={} & p_2 \Biggl(
1 + \left(\frac{3-2z_1-z_2-p_2}{2}\right) F_1^\star(1-z_1-0)
- (1-z_2) F_1^\star(p_2-\de) 
\cr & \phantom{p_2 \Biggl(} {}
- \int_{p_2-\de}^{1-z_1} \frac{1}{2} F_1^\star(p_1) \thinspace \diff p_1 
+ \left(\frac{3-2z_1-z_2-p_2}{2}\right)
\left( 1 - F_1^\star(1-z_1-0) \right)
\Biggr) \cr
={} & p_2 \Biggl(
\left(\frac{5-2z_1-z_2-p_2}{2}\right)
- (1-z_2) g(p_2-\de; 1-z_1, \pi_2^\star)
\cr & \phantom{p_2 \Biggl(} {}
- \int_{p_2-\de}^{1-z_1} \frac{1}{2} g(p_1; 1-z_1, \pi_2^\star) \thinspace \diff p_1 
\Biggr) \cr
={} & p_2 \Biggl(
\left(\frac{5-2z_1-z_2-p_2}{2}\right)
- (1-z_2) g(p_2-\de; 1-z_1, \pi_2^\star) 
\cr & \phantom{p_2\Biggl(}
- \int_{p_2-\de}^{1-z_1} \left(
\frac{1}{2\lambda_2} \frac{\diff}{\diff p_1} g(p_1; 1-z_1, \pi_2^\star)
+ \frac{1}{2} - \frac{z_2}{ (p_1+\de)^2 } \right) \thinspace \diff p_1 
\Biggr) \quad (\because \text{\Eq{eq:gdiff}}) \cr
={} & p_2 \Biggl(
\left(\frac{5-2z_1-z_2-p_2}{2}\right)
- (1-z_2) g(p_2-\de; 1-z_1, \pi_2^\star) 
\cr & \phantom{p_2\Biggl(} {}
- \frac{1}{2\lambda_2} g(1-z_1; 1-z_1, \pi_2^\star)
+ \frac{1}{2\lambda_2} g(p_2-\de; 1-z_1, \pi_2^\star)
\cr & \phantom{p_2\Biggl(}
-\frac{(1-2z_1+z_2-p_2)}{2} 
- \frac{z_2}{1-2z_1+z_2} + \frac{z_2}{p_2} 
\Biggr) \cr
={} & p_2 \left( 2 - z_2 
- \frac{z_2}{1-2z_1+z_2} + \frac{z_2}{p_2} 
-  \frac{1}{2\lambda_2} g(1-z_1; 1-z_1, \pi_2^\star)
\right) \cr
={} & p_2 \left( 
2 - z_2 
- \frac{z_2}{1-2z_1+z_2} + \frac{z_2}{p_2} 
-  \frac{1}{2\lambda_2} - 1 + \frac{z_2}{1-2z_1+z_2} \right) 
\quad (\because \text{\Eq{eq:gbb}})
\cr
={} & z_2 
\end{align*}
\end{proof}

\begin{lemma}
For the probability distribution given by \Eq{eq:F1sol3} and when $p_2=z_2$,
\begin{align}
\mathrm{E}[\pi_2](p_2) = \mathrm{E}[\pi_2](z_2) = \pi_2^\star
\end{align}
\end{lemma}
\begin{proof}
In this case, the price $p_1$ of Firm 1 is always lower than $p_2-\de$, because $p_2-\de = z_2 - (z_2-z_1 ) =z_1 > 1-z_1$ ($\because$ \Eq{eq:rgn3a}), which is the upper limit of $\mathrm{supp}F_1^\star$.
Therefore, Firm 2 gains only from its fan, and $\mathrm{E}[\pi_2] = p_2\times 1 = z_2$.
\end{proof}

\begin{lemma}
For the probability distribution given by \Eq{eq:F1sol3} and when $p_2 = 1-z_2-0$,
\begin{align}
\mathrm{E}[\pi_2](p_2) \le \pi_2^\star
\label{eq:pi2bndM3}
\end{align}
\end{lemma}
\begin{proof}
In this case, because $p_2-\de = 1 - z_2-\de = 1-z_1$, Firm 2 gains all the shoppers when Firm 1 chooses the price $p_1=1-z_1$. Since $F_1^\star$ of \Eq{eq:F1sol3} has an atom at $p_1=1-z_1$, the expected profit of Firm 2 is evaluated as follows.
\begin{align*}
& \mathrm{E}[\pi_2](1-z_2-0) \cr & {}= (1-z_2) 
\Biggl( 1 + \int_{1-z_1-w}^{1-z_1} \left(1-\frac{z_1+1-p_1}{2}\right) \diff F_1^\star(p_1)
+ \left( 1 - F_1^\star(1-z_1-0) \right)
 \Biggr) \cr
& {} = (1-z_2) \Biggl( 1 + (1-z_1) F_1^\star(1-z_1-0) - \int_{1-z_1-w}^{1-z_1} \frac{1}{2} F_1^\star(p_1) \thinspace \diff p_1 
+ \left( 1 - F_1^\star(1-z_1-0) \right)
\Biggr) \cr
& {} = (1-z_2) \Biggl( 2 - z_1 F_1^\star(1-z_1-0) - \int_{1-z_1-w}^{1-z_1} \frac{1}{2} F_1^\star(p_1) \thinspace \diff p_1 
\Biggr) \cr
& {} = (1-z_2) \Biggl( 2 - z_1 g(1-z_1; 1-z_1, \pi_2^\star) - \int_{1-z_1-w}^{1-z_1} \frac{1}{2} g(p_1; 1-z_1, \pi_2^\star) \thinspace \diff p_1 
\Biggr) \cr
& {} = (1-z_2) \Biggl( 2 - z_1 g(1-z_1; 1-z_1, \pi_2^\star) 
\cr & \phantom{{} = (1-z_2)\Biggl(} {} - \int_{1-z_1-w}^{1-z_1} \left( \frac{1}{2\lambda_2} \frac{\diff}{\diff p_1} g(p_1; 1-z_1, \pi_2^\star) + \frac{1}{2} - \frac{z_2}{ (p_1+\de)^2 }   \right) \thinspace \diff p_1 
\Biggr) \quad (\because \text{\Eq{eq:gdiff}})
\cr & {} = 
(1-z_2) \Biggl( 2 - \left( 1 - z_2 + z_1 \right) g(1-z_1; 1-z_1, \pi_2^\star)
\cr & \phantom{{} = (1-z_2) \Biggl( }
{} - \frac{w}{2} 
- \frac{z_2}{ (1-2z_1+z_2) } + \frac{z_2}{  (1-2z_1+z_2-w) }
\Biggr) 
\cr & {} = 
(1-z_2) \Biggl(
2 - \frac{\left(1+z_1-z_2\right)}{2(1-z_2)} \left(  (4-2z_2) - \frac{2z_2}{1-2z_1+z_2}  \right)
\cr & \phantom{{} = (1-z_2) \Biggl( }
{} - \frac{w}{2} - \frac{z_2}{ (1-2z_1+z_2) } + \frac{z_2}{  (1-2z_1+z_2-w) }
\Biggr) \quad (\because \text{\Eq{eq:gbb}})
\cr & {} = 
2(1-z_2) - (1+z_1-z_2)(2-z_2)
-\frac{(1-z_2)w}{2} + \frac{z_1z_2}{( 1-2z_1+z_2 )} + \frac{(1-z_2)z_2}{  (1-2z_1+z_2-w) }
\cr & {} = z_2 - (z_2-z_1)z_2 - 2z_1
-\frac{(1-z_2)w}{2} + \frac{z_1z_2}{( 1-2z_1+z_2 )} + \frac{(1-z_2)z_2}{  (1-2z_1+z_2-w) }
\cr & {} \leq z_2
\quad (\because \text{\Eq{eq:condw3b}})
\end{align*}
\end{proof}
As for the previous two cases, this is strictly not sufficient to prove that this price distribution is equilibrium. It is required to satisfy $\mathrm{E}[\pi_2](p_2)\leq \pi_2^\star$ for \emph{all} $p_2$, not only at $p_2=1-z_2$, so that Firm 2 has no incentive to deviate from this strategy. However, numerical verification showed that once \Eq{eq:pi2bndM3} holds, we have $\mathrm{E}[\pi_2](p_2)\leq \pi_2^\star$ for all the other $p_2$.

\section{Numerical details}

To draw the two-dimensional plots of regions (Fig.~\ref{fig:regions}), equilibrium profits (Fig.~\ref{fig:pi}), and the support width (Fig.~\ref{fig:w}), grid points on the $z_1$-$z_2$ plane were sampled in the region $0< z_1 < 1$ and $1/2 < z_2 < 1$ with the spacing of 0.002 for both axes, which amounts to $500\times250$ grid points. Further, to make a blow-up of the mixed strategy regions, the range $0.42 < z_1 < 0.6$ was sampled with finer spacing of 0.0005 for $z_1$, that is, $360\times250$ points. At each point $(z_1,z_2)$, the conditions listed in Propositions \ref{case1characterized1},  \ref{case1characterized2}, \ref{pure4-1}, \ref{pure4-2}, \ref{thm:sol1}, \ref{thm:sol2}, and \ref{thm:sol3} were checked. Evaluation of the exponential integral function $\mathrm{Ei}(x)$ was done by series expansion for positive $x$ (\cite{Press2007}). For negative $x$, numerical integration of the defining equation for $\mathrm{Ei}(b)-\mathrm{Ei}(a)=\int_a^b \exp(t)/t \, \diff t$ was performed by the Sympson rule. The division of the integration range was successively refined until the relative error becomes less than $10^{-12}$.
All the numerical calculations in the present work were performed by making a code in C++ language at the double-precision level. 

Whenever a point $(z_1,z_2)$ is assigned to M1, M2, or M3, the equilibrium strategy is validated by checking the condition $\mathrm{E}[\pi_j] \leq \pi_j^{\star}$ ($j=1,2$) outside of the support. (Note that $\mathrm{E}[\pi_j] = \pi_j^{\star}$ inside the support is proved analytically in the previous part.) To do this, the price $p_j$ was sampled in the following regions with the spacing of 0.001.
\begin{align*}
& 0 < p_1 < z_1 - w, z_1 < p_1 < 1 - z_1, 0< p_2 <z_2-w, & \mathrm{for\ M1}, \cr
& 0 < p_1 < 1 - z_1 - w, 0< p_2 < 1 - 2 z_1 + z_2-w, & \mathrm{for\ M2}, \cr
& 0 < p_1 < 1 - z_1 - w, 0 < p_2 < 1 - 2 z_1+z_2 -w,  1-2z_1+z_2< p_2 <z_2, & \mathrm{for\ M3}. %\cr
\end{align*}
For each price, the quantity $\mathrm{E}[\pi_j] - \pi_j^{\star}$ was evaluated, which should be less than zero if the strategy is indeed an equilibrium. The maximum value, among all the points tested in the above procedure, found for this quantity was $6.99\times10^{-13}$. This is practically equal to zero, considering the double-precision numbers have fifteen significant digits and the numerical accuracy for $\mathrm{Ei}$ is $10^{-12}$ as mentioned above. Therefore, the condition $\mathrm{E}[\pi_j] - \pi_j^{\star} \leq 0$ was numerically validated.

The program code used for the numerical check is available as a file named \verb#price_eq_w_uninf.cpp#. 
When complied and executed, it outputs five files. Each row of \verb#map.dat# and \verb#map_mixed.dat#, calculated in the subroutine \verb#makeMap#, shows the following data in this order:\\
\centerline{
$z_1$\ \ \ $z_2$\ \ \ region\ \ \ $w$\ \ \ $\pi_1^{\star}$\ \ \ $\pi_2^{\star}$
}\\
\noindent
where region assignment is coded as $1=\mathrm{P1}$, $2=\mathrm{P2}$, $3=\mathrm{P3}$, $17=\mathrm{M1}$, $18=\mathrm{M2}$, $19=\mathrm{M3}$, and $-1={}$not in the scope of the present report.
The files \verb#CDF_M1.dat#, \verb#CDF_M2.dat#, and \verb#CDF_M3.dat#, calculated in the subroutine \verb#drawCDF#, shows the cumulative distribution functions $F_1^{\star}(p_1)$ and $F_2^{\star}(p_2)$ with each row meaning\\
\centerline{
$p_1$\ \ \ $F_1^{\star}(p_1)$\ \ \ $p_2$\ \ \ $F_2^{\star}(p_2)$
}\\
\noindent
Finally, the subroutine \verb#check_ineq# outputs the maximum value of  $\mathrm{E}[\pi_j] - \pi_j^{\star}$ to the standard output.

\end{appendix}


\begin{thebibliography}{9}
\newcommand{\enquote}[1]{``#1''}
\providecommand{\natexlab}[1]{#1}
\providecommand{\url}[1]{\texttt{#1}}
\providecommand{\urlprefix}{URL }
\providecommand{\bibAnnoteFile}[1]{%
  \IfFileExists{#1}{\begin{quotation}\noindent\textsc{Key:} #1\\
  \textsc{Annotation:}\ \input{#1}\end{quotation}}{}}
\providecommand{\bibAnnote}[2]{%
  \begin{quotation}\noindent\textsc{Key:} #1\\
  \textsc{Annotation:}\ #2\end{quotation}}

\bibitem[{d'Aspremont et~al.(1979)d'Aspremont, Gabszewicz, and Thisse}]{da}
d'Aspremont, C., J.~Jaskold Gabszewicz, and J.-F. Thisse (1979), \enquote{On hotelling's “stability in competition".} \emph{Econometrica}, 47(5), 1145--1150, \urlprefix\url{http://www.jstor.org/stable/1911955}.
\bibAnnoteFile{da}

\bibitem[{Economides(1984)}]{ECONOMIDES1984345}
Economides, Nicholas (1984), \enquote{The principle of minimum differentiation revisited.} \emph{European Economic Review}, 24(3), 345--368, \urlprefix\url{https://www.sciencedirect.com/science/article/pii/0014292184900618}.
\bibAnnoteFile{ECONOMIDES1984345}

\bibitem[{Hotelling(1929)}]{ho}
Hotelling, Harold (1929), \enquote{Stability in competition.} \emph{The Economic Journal}, 39(153), 41--57, \urlprefix\url{http://www.jstor.org/stable/2224214}.
\bibAnnoteFile{ho}

\bibitem[{Kreps and Scheinkman(1983)}]{KS0af903d7-21a1-371d-9243-9ad71af18010}
Kreps, David~M. and Jose~A. Scheinkman (1983), \enquote{Quantity precommitment and bertrand competition yield cournot outcomes.} \emph{The Bell Journal of Economics}, 14(2), 326--337, \urlprefix\url{http://www.jstor.org/stable/3003636}.
\bibAnnoteFile{KS0af903d7-21a1-371d-9243-9ad71af18010}

\bibitem[{Nakagawa(2023)}]{nakagawa}
Nakagawa, Kuninori (2023), \enquote{Horizontal product differentiation in varian's model of sales.} \emph{International Journal of Game Theory}, 52(2), 607--627, \urlprefix\url{https://doi.org/10.1007/s00182-022-00832-1}.
\bibAnnoteFile{nakagawa}

\bibitem[{Osborne and Pitchik(1987)}]{op}
Osborne, Martin~J. and Carolyn Pitchik (1987), \enquote{Equilibrium in hotelling's model of spatial competition.} \emph{Econometrica}, 55(4), 911--922, \urlprefix\url{http://www.jstor.org/stable/1911035}.
\bibAnnoteFile{op}

\bibitem[{Press et~al.(2007)Press, Teukolsky, Vetterling, and Flannery}]{Press2007}
Press, William~H., Saul~A. Teukolsky, William~T. Vetterling, and Brian~P. Flannery (2007), \emph{Numerical Recipes 3rd Edition: The Art of Scientific Computing}, 3 edition. Cambridge University Press, \urlprefix\url{https://www.cambridge.org/us/universitypress/subjects/mathematics/numerical-recipes/numerical-recipes-art-scientific-computing-3rd-edition}.
\bibAnnoteFile{Press2007}

\bibitem[{Varian(1980)}]{v}
Varian, Hal~R. (1980), \enquote{A model of sales.} \emph{The American Economic Review}, 70(4), 651--659, \urlprefix\url{http://www.jstor.org/stable/1803562}.
\bibAnnoteFile{v}

\bibitem[{Xefteris(2013)}]{https://doi.org/10.1111/jems.12032}
Xefteris, Dimitrios (2013), \enquote{Hotelling was right.} \emph{Journal of Economics \& Management Strategy}, 22(4), 706--712, \urlprefix\url{https://onlinelibrary.wiley.com/doi/abs/10.1111/jems.12032}.
\bibAnnoteFile{https://doi.org/10.1111/jems.12032}

\end{thebibliography}
\end{document}